\documentclass[a4paper,fleqn]{cas-sc}

\AddToHook{begindocument/before}{\ifdefined\bibsep\else\newlength\bibsep\fi} % class is broken...

\ExplSyntaxOn
\exp_args:NNno \exp_args:Nno \use:n { \cs_gset:Npn \__make_fig_caption:nn #1#2 }
  {
    \exp_after:wN \use_ii_i:nn \exp_after:wN
      { \__make_fig_caption:nn {#1} {#2} }
      { \dim_set:Nn \l_fig_width_dim \linewidth }
  }
\exp_args:NNno \exp_args:Nno \use:n { \cs_gset:Npn \__make_tbl_caption:nn #1#2 }
  {
    \exp_after:wN \use_ii_i:nn \exp_after:wN
      { \__make_tbl_caption:nn {#1} {#2} }
      { \dim_set:Nn \l_tbl_width_dim \linewidth }
  }
\ExplSyntaxOff

\usepackage[numbers]{natbib}

\usepackage{hyperref}
\usepackage{amssymb}
\usepackage{amsmath}
\usepackage{tablefootnote}
\usepackage{color}
\usepackage{graphicx}   % For \resizebox
\usepackage{makecell}   % For \makecell (multi-line cells)
\usepackage{booktabs}   % For \toprule, 
\usepackage{array}
\newcolumntype{P}[1]{>{\raggedleft\arraybackslash}p{#1}}

\usepackage{multirow}% multirow for diagonal box
\usepackage{subcaption}
\usepackage{adjustbox}
\usepackage{float}
\usepackage[frozencache,cachedir=mintedcache]{minted}
\usepackage{multicol}

\begin{document}
\let\WriteBookmarks\relax
\let\printorcid\relax
\def\floatpagepagefraction{1}
\def\textpagefraction{.001}

% \shorttitle{GNN-ACLP: Graph Neural Networks Based Analog Circuit Link Prediction}    
\shorttitle{Graph Neural Networks Based Analog Circuit Link Prediction}    

\shortauthors{Pan et al.}  

\author[a]{Guanyuan Pan}
 \ead{guanyuanpeterpan@gmail.com}
 \credit{Writing – original draft, Visualization, Validation, Software, Investigation}
 
%\QUERY[2]
\author[b]{Tiansheng Zhou}
 \ead{zhoutiansheng_2024@163.com}
 \credit{Visualization, Data curation}
%\QUERY[3]
\author[b]{Jianxiang Zhao}
 \ead{246270147@hdu.edu.cn}
 \credit{Writing – review \& editing}
 
\author[b]{Zhi Li}
 \ead{lizhi2513@hdu.edu.cn}
 \credit{Visualization}
 
\author[d]{Yugui Lin}
 \ead{yuguilin0209@163.com}
 \credit{Validation}
 
\author[b]{Bingtao Ma}
 \cormark[1]
% \cortext[1]{\CORS
% \TFadd[\AffNum].
% }
\ead{mabingtao93@gmail.com}
\credit{Writing – review \& editing, Project administration, Supervision, Writing – original draft}

\author[c]{Yaqi Wang}
 \cormark[1]
 \ead{wangyaqi@hdu.edu.cn}
 \credit{Writing – review \& editing, Conceptualization, Supervision}
 
\author[e]{Pietro Li\`o}
 \cormark[1]
 \ead{pl219@cam.ac.uk}
 \credit{Writing – review \& editing, Supervision}
 
\author[b]{Shuai Wang}
\cormark[1]
\ead{shuaiwang.tai@gmail.com}
\credit{Writing – review \& editing, Supervision, Resources}

\cortext[CorrespondingAuthor]{Corresponding authors.}

\affiliation[a]{o={HDU-ITMO Joint Institute, Hangzhou Dianzi University}, a={No. 1158, 2nd Avenue, Xiasha Higher Education Zone, Jianggan District}, c={Hangzhou}, p={310018}, s={Zhejiang Province}, cy={China}}

\affiliation[b]{o={Intelligent Information Processing Laboratory, Hangzhou Dianzi University}, a={No. 1158, 2nd Avenue, Xiasha Higher Education Zone, Jianggan District}, c={Hangzhou}, p={310018}, s={Zhejiang Province}, cy={China}}

% \affiliation[c]{o={College of Media Engineering, Communication University of Zhejiang}, a={No. 998, Xueyuan Street, Xiasha Higher Education Zone, Jianggan District}, c={Hangzhou}, p={310018}, s={Zhejiang Province}, cy={China}}

\affiliation[c]{o={Innovation Center for Electronic Design Automation Technology, Hangzhou Dianzi University}, a={No. 1158, 2nd Avenue, Xiasha Higher Education Zone, Jianggan District}, c={Hangzhou}, p={310018}, s={Zhejiang Province}, cy={China}}

\affiliation[d]{o={School of Computer Science and Technology, South China Business College, Guangdong University of Foreign Studies}, a={181 Liangtian Middle Road, Baiyun District}, c={Guangzhou}, p={510545}, s={Guangdong Province}, cy={China}}

\affiliation[e]{o={Department of Computer Science and Technology, University of Cambridge, William Gates Building}, a={15 JJ Thomson Avenue, Cambridge, CB3 0FD, Cambridgeshire}, c={England}, cy={United Kingdom}}

% \title [mode = title]{GNN-ACLP: Graph Neural Networks Based Analog Circuit Link Prediction}  
\title [mode = title]{Graph Neural Networks Based Analog Circuit Link Prediction}  

%% Abstract
\begin{abstract}
Circuit link prediction, which identifies missing component connections from incomplete netlists, is crucial in analog circuit design automation. However, existing methods face three main challenges: 1) Insufficient use of topological patterns in circuit graphs reduces prediction accuracy; 2) Data scarcity due to the complexity of annotations hinders model generalization; 3) Limited adaptability to various netlist formats restricts model flexibility. We propose Graph Neural Networks Based Analog Circuit Link Prediction (GNN-ACLP), a graph neural networks (GNNs) based method featuring three innovations to tackle these challenges. First, we introduce the SEAL (learning from Subgraphs, Embeddings, and Attributes for Link prediction) framework and achieve port-level accuracy in circuit link prediction. Second, we propose Netlist Babel Fish, a netlist format conversion tool that leverages retrieval-augmented generation (RAG) with a large language model (LLM) to enhance the compatibility of netlist formats. Finally, we build a comprehensive dataset, SpiceNetlist, comprising 775 annotated circuits of 7 different types across 10 component classes. Experiments demonstrate accuracy improvements of 16.08\% on SpiceNetlist, 11.38\% on Image2Net, and 16.01\% on Masala-CHAI compared to the baseline in intra-dataset evaluation, while maintaining accuracy from 92.05\% to 99.07\% in cross-dataset evaluation, demonstrating robust feature transfer capabilities. However, its linear computational complexity makes processing large-scale netlists challenging and requires future addressing.
\end{abstract}

%\nocite{*}

% Keywords

\maketitle

\section{Introduction}

Analog circuit design has mainly been performed manually by experienced engineers \cite{krylov2023learning}. However, analog design is susceptible to layout geometry \cite{scheible2015automation}, and even experienced engineers face difficulties when designing them. Moreover, analog circuits are both heterogeneous and hierarchical, leading to a variety of representations. For example, a differential pair can be represented by two transistors or by eight segmented devices; a current mirror might cascade through three levels or fold into a single branch. When converting to graphs, these different representations result in graphs with varying vertex counts and degrees, even though the circuits are electrically equivalent.

Additionally, there is no definitive criterion for labeling functional boundaries. Consequently, it is equally valid to embed a bias network within an operational transconductance amplifier (OTA) symbol or to keep it separate, resulting in two incompatible topologies representing the same circuit. As a result, existing automated solutions that optimize circuit parameters for a specific design specification \cite{zhang2025mce} are often inaccurate, schematic-only, and lack generalizability \cite{settaluri2020autockt}. Therefore, analog circuit design automation has recently gained significant attention \cite{Sturm_2019}. 

One main task of analog design automation is circuit link prediction, which infers missing component interconnections from incomplete netlists. There are three conventional methods for general link prediction: heuristic methods, latent feature methods, and learning-based methods. Although heuristic methods offer simplicity and interpretability through predefined topological features, their generalizability is limited by dependence on predefined similarity metrics that encode fixed assumptions about link formation, leading to their inability to adapt to the structural heterogeneity of diverse networks \cite{liben2003link, haghani2019systemic}. Therefore, researchers do not commonly use them in circuit link prediction. Latent-feature methods compute latent properties or node representations, typically through factorizing a specific matrix derived from the network. Although effective, the generalizability of latent-feature methods is often constrained by their reliance on specific factorization assumptions and techniques \cite{haghani2019systemic}.

Therefore, in the Electronic Design Automation (EDA) field, where circuit netlists represent graph-structured data, machine learning (ML) methods have gained more prominence. Genssler et al. proposed a novel method using brain-inspired hyperdimensional computing (HDC) for encoding and recognizing gate-level netlists \cite{genssler_hdcircuit_2024}. However, the high-dimensional vector operations involved in HDC led to significant computational and storage overhead, especially when dealing with large-scale graph data \cite{Ge_2020}. Luo et al. proposed the Directional Equivariant Hypergraph Neural Network (DE-HNN) for effective representation learning on directed hypergraphs, particularly for circuit netlists in chip design \cite{luo_-hnn_2024}. However, DE-HNN's hypergraph diffusion and tensor operations may result in exponentially increasing computational costs as the hyperedge order increases, making it challenging to scale for large chip design \cite{kim2022equivarianthypergraphneuralnetworks, choe2023classification, ruggeri2024message, neuhauser2024learning}. 

GNNs can leverage the inherent graph structure of circuit netlists, making them well-suited for circuit link prediction. However, a critical challenge is the lack of annotated training data due to high human workload and data sensitivity \cite{jiang2023circuitnet}. Furthermore, the diversity of netlist formats poses significant challenges in analog design automation. In addition, engineers often create customized netlist formats by modifying standard formats \cite{patilesim}. Consequently, different datasets often exist in various formats, which complicates the training processes for machine learning methods.

\begin{figure}
\centering
\includegraphics[width=0.92\linewidth]{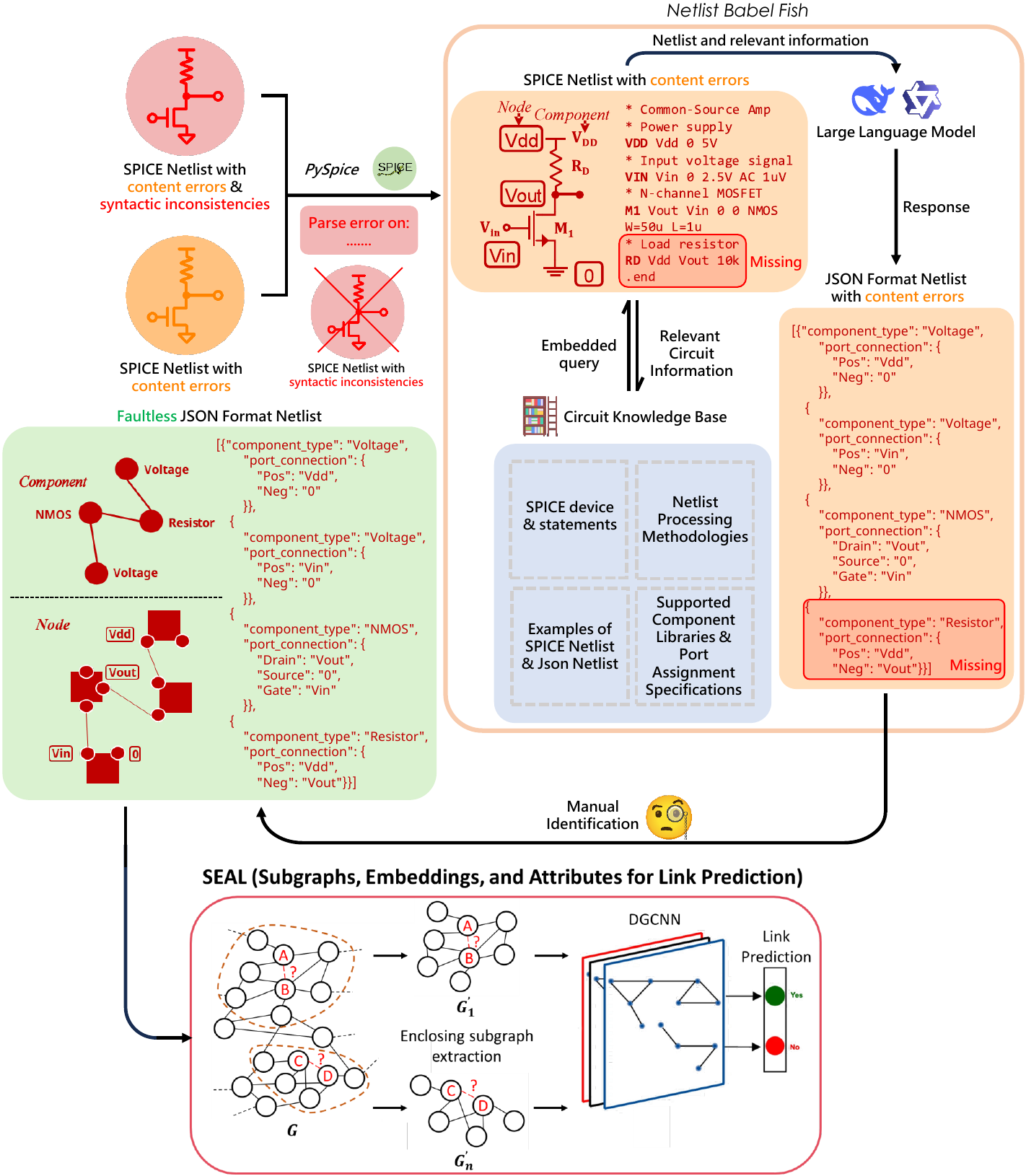}
\caption{GNN-ACLP overview. We first preprocess SPICE netlist datasets by correcting errors and syntactic inconsistencies, generating faultless JSON netlists. We then enable bidirectional format conversion via Netlist Babel Fish, a RAG-based LLM framework plus domain-specific knowledge. We finally model the netlists as graphs and apply SEAL for circuit link prediction.}
\label{GNN-ACLP}
\end{figure}

To overcome these obstacles, we propose \textbf{GNN-ACLP}, a GNN-based method for analog circuit link prediction, as demonstrated in Figure \ref{GNN-ACLP}. \textbf{1)} We formulate circuit link prediction as an undirected graph link prediction problem, where nodes represent circuit components and edges denote connections. To solve this task, we leverage the SEAL framework, trained using Double-Radius Node Labeling (DRNL) one-hot encoding augmented with native component node characteristics for enhanced representation learning. \textbf{2)} We introduce \textbf{Netlist Babel Fish}, a framework featuring RAG for LLM-based interpretation and enabling bidirectional netlist format conversion through domain-specific knowledge integration, to overcome the challenge of differing dataset formats. \textbf{3)} We develop a robust preprocessing pipeline to detect and correct content errors and syntactic inconsistencies across SPICE netlist datasets. We then apply the pipeline to two small-scale datasets and integrate them into a single, unified large-scale dataset, \textbf{SpiceNetlist}, to support the training and evaluation of circuit link prediction methods. SpiceNetlist has 775 annotated circuits in both SPICE and JSON formats, featuring 10 component types. We present its statistics in Table \ref{spicenetlist}. The experimental results demonstrate significant performance gains in intra-dataset evaluation, achieving accuracy improvements of 16.08\% on SpiceNetlist, 11.38\% on Image2Net and 16.01\% on Masala-CHAI compared to the baseline. Furthermore, our method maintains accuracy from 92.05\% to 99.07\% in cross-dataset evaluation, demonstrating robust feature transfer capabilities.

\begin{table}
\centering
\caption{SpiceNetlist statistics.}
\resizebox{0.5\textwidth}{!}{
\begin{tabular}{lccccc}
\hline
\textbf{Dataset} & \textbf{Graphs} & \textbf{Classes} & \textbf{Avg. Nodes} & \textbf{Avg. Edges} \\
\hline
SpiceNetlist & 775 & 10 & 13.78 & 15.60 \\
\hline
\end{tabular}
}
\label{spicenetlist}
\end{table}

The contributions of our work are outlined as follows:

\begin{itemize}

\item \textbf{GNN-ACLP}, a port-level accurate circuit link prediction method based on the SEAL framework using GNNs. To the best of our knowledge, no previous research has investigated circuit link prediction at the port level, unlike existing component-level methods.

\item \textbf{Netlist Babel Fish}, a netlist format converter enabling port-level netlist parsing across multiple formats, leveraging RAG for LLM-based interpretation.

\item \textbf{SpiceNetlist}, a novel circuit netlist dataset comprising 775 annotated circuits in both SPICE and JSON formats, featuring 10 component types to facilitate training and evaluation of circuit link prediction methods.

\end{itemize}

\section{Preliminaries and problem formulation}

The circuit link prediction problem is a part of the circuit completion problem. We can define the circuit completion problem as follows:

\textbf{Problem 2.1} (circuit completion problem) Let $ v_0 $ and $ v_1 $ represent the netlists of two circuit schematics, where $v_1$ is a partial copy of $v_0$ missing all connections associated with a component $x$ that exist in $v_0$. The circuit completion problem involves predicting the category of the missing component $x$ and all of its missing connections, where 

\begin{equation} \label{eq:component_set}
   x \in \{1, 2, 3, \ldots, k\}. 
\end{equation}

Here, netlists define the connections between circuit components, including ports, each with a unique ID and type. 

There are several challenges in solving circuit completion problems. First, the problem implicitly relies on a clear standard to determine a circuit’s validity, which is necessary for verifying the correctness of a solution. Although we can perform some checks, no definitive criteria exist to validate a circuit netlist. Second, it may not be relevant to our interests even if a circuit is valid. As long as both $v_{0}$ and $v’_{0}$ are netlists of designs that are considered “interesting” (where “interesting” is defined by the particular use case), either could potentially serve as a valid completion for the partial netlist $v_{1}$. However, $v’_{0}$ may be a trivial completion irrelevant to the particular application being studied \cite{6782689, 6580963}. To address these challenges, we discard any supplementary details from the netlists and convert each netlist into an undirected graph, denoted as 

\begin{equation} \label{eq:graph}
    G = (V,E,\phi ).
\end{equation}

In this graph, the set of vertices $V$ corresponds to the ports of all components in the netlist, while the edges denote the connections between these ports in $E$ that link the corresponding vertices. Additionally, each vertex $v$ has an integer type, denoted as $\phi (v)$, taken from the set defined in \eqref{eq:component_set}.

It is worth mentioning that presenting netlists in graphs at the port level, rather than the component level, is more suitable for circuit link prediction. Analog circuit components have multiple ports, each with a distinct function. Modeling components as single nodes would result in the loss of critical pin constraints, potentially leading to non-unique or invalid prediction outcomes. By modeling at the port level, we represent each pin separately and ensures the legality of connections \cite{xu2024graph}. Furthermore, we indicate the relative distances and roles of ports by DRNL, allowing the model to learn common circuit topology patterns implicitly, while component-level models can only depict the simple structure of two connected components.

We subsequently formulate the circuit link prediction problem in this manner:

\textbf{Problem 2.2} (circuit link completion problem on graphs)\label{Problem 2.2} Consider a graph $G'$ created by deleting a random vertex $u\in V$ from $G$. Using the value of $\phi (u)$ and $G'$, determine all vertices adjacent to $u$ in $G$.

We define the vertex adjacency for node $u$ as:
\begin{equation} \label{eq:neighbor_node_set}
    \{v\in V: (u,v)\in E\}.
\end{equation}

\begin{figure}
  \centering
  \begin{minipage}{0.48\textwidth}
    \centering
    \begin{subfigure}{\linewidth}
      \centering
      \includegraphics[width=\linewidth]{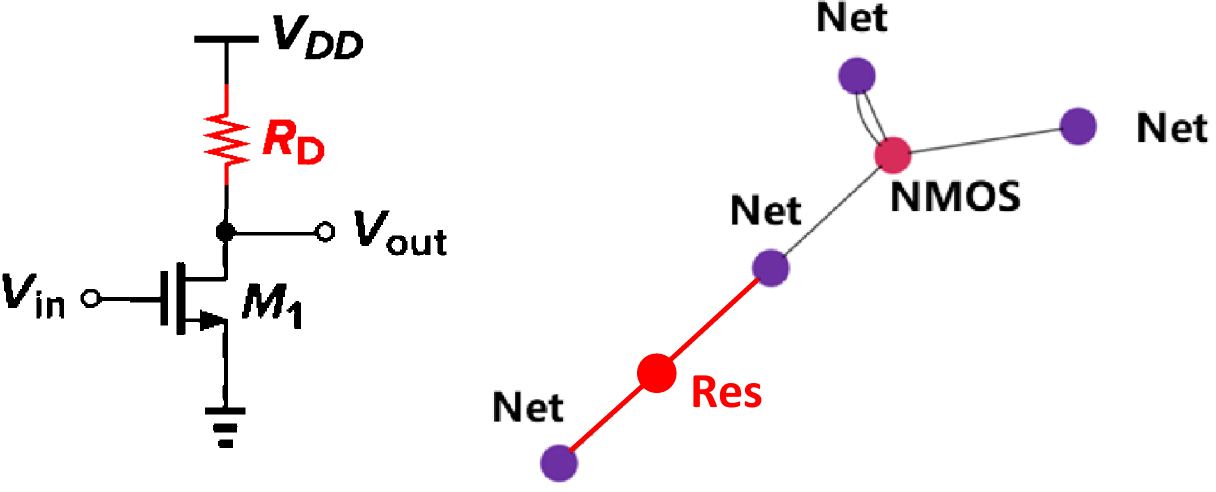}
      \caption{Schematic representation of a netlist.}
    \end{subfigure}
    \par\bigskip
    \begin{subfigure}{\linewidth}
      \centering
      \includegraphics[width=\linewidth]{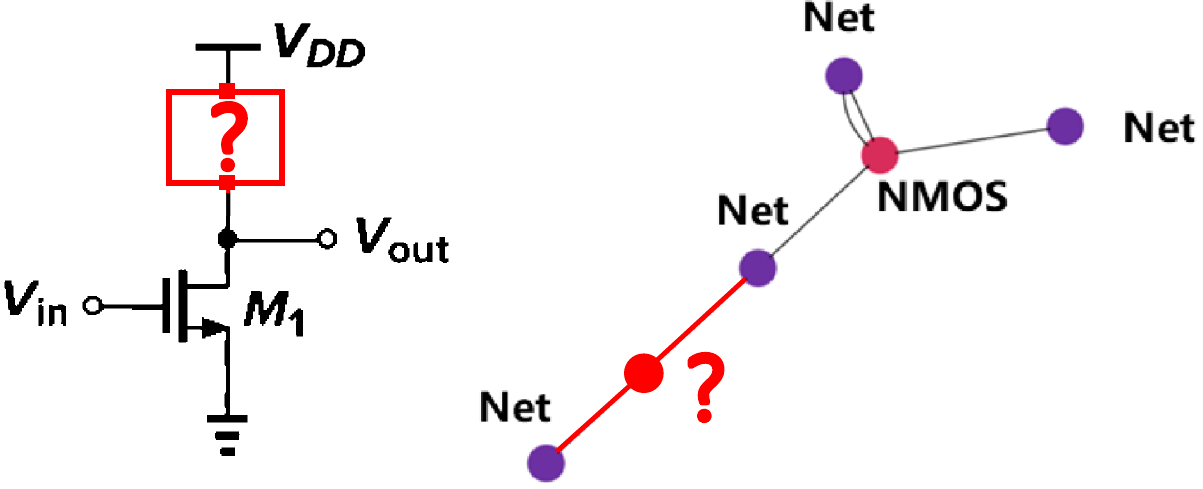}
      \caption{Schematic representation of a netlist with a missing component.}
    \end{subfigure}
    \par\medskip
    \caption{Visualization of the circuit link prediction problem.}
    \label{Illustrationofthelinksinsertionproblem}
  \end{minipage}
  \hfill
  \begin{minipage}{0.48\textwidth}
    \centering
    \includegraphics[width=\linewidth]{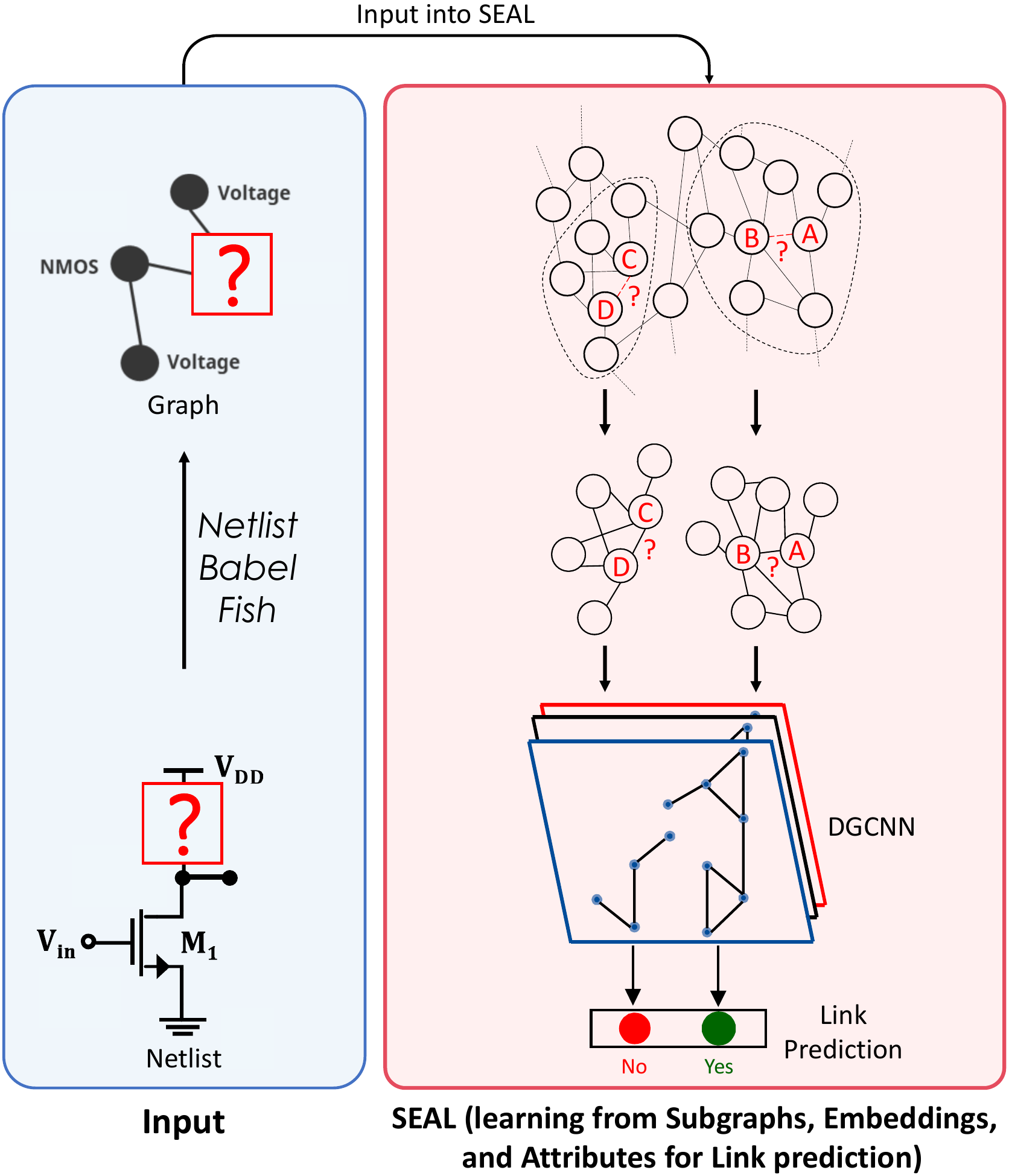}
    \caption{The architecture of our circuit link prediction architecture with SEAL. Netlists are converted into graph representations and processed by SEAL, which extracts and encloses subgraphs to generate training samples for GNN-based link prediction. We implement DGCNN as the GNN~architecture.}
    \label{architecture_SEAL_link_prediction}
  \end{minipage}
\end{figure}

We propose data-driven solutions to tackle the problem. Specifically, we learn a map:

\begin{equation} \label{eq:map}
    {\hat{\xi }}: {G},[k]\rightarrow 2^V,
\end{equation}

where $G$ is an input graph and $\phi \in [k]$ denotes a component type with $k$ possibilities. This mapping returns a subset of vertices (the power set of vertices $V$, represented as $2^V$) that indicate the connections for the ports of components within $G$. We provide visualizations of this problem in Figure \ref{Illustrationofthelinksinsertionproblem}.

\section{Link prediction through GNNs}
\label{4.2}
\hyperref[Problem 2.2]{Problem 2.2} presents a graph $G$ that contains various node types and highlights a distinguished node $v$. The problem aims to predict the neighbors of $ v $ within $ G $. We can simplify this problem by determining whether an edge should exist between $ v $ and each neighboring node $ u $ in $ G $. The aim is to evaluate whether two nodes should be connected, depending on the existing linkage patterns in the graph. Therefore, we can frame this connection test as a link prediction~problem.

We select the SEAL framework \cite{zhang2018link}, which we visualize in Figure \ref{architecture_SEAL_link_prediction}. SEAL captures subgraphs surrounding connected node pairs from positive examples and extracts training data for these pairs. For a pair of nodes $(x, y)$, an enclosing subgraph is defined as a subgraph that includes the h-hop neighborhood of both $x$ and $y$. SEAL also employs a node labeling approach, DRNL, to label nodes within the subgraphs, which aims to identify the different roles of nodes while preserving structural information. We adopt the SEAL framework for link prediction in circuit graphs and train it using DRNL one-hot encoding combined with the native component node characteristics.

It is worth mentioning that our work may not address the circuit link prediction problem on graphs since there can be multiple valid ways to link ports in a partial circuit \cite{hibshman2023inherentlimitstopologybasedlink, chen2021topology}. Therefore, the proposed method may occasionally encounter failures. However, we demonstrate through extensive experiments that graph machine-learning methods yield reliable results on real-world datasets, supporting human experts in the design, synthesis, and evaluation~processes.

\section{Netlist Babel Fish: Netlist Format Converter}

\begin{figure*}%% placement specifier
%% Use \includegraphics command to insert graphic files. Place graphics files in 
%% working directory.
\centering%% For centre alignment of image.
\includegraphics[width=0.92\linewidth]{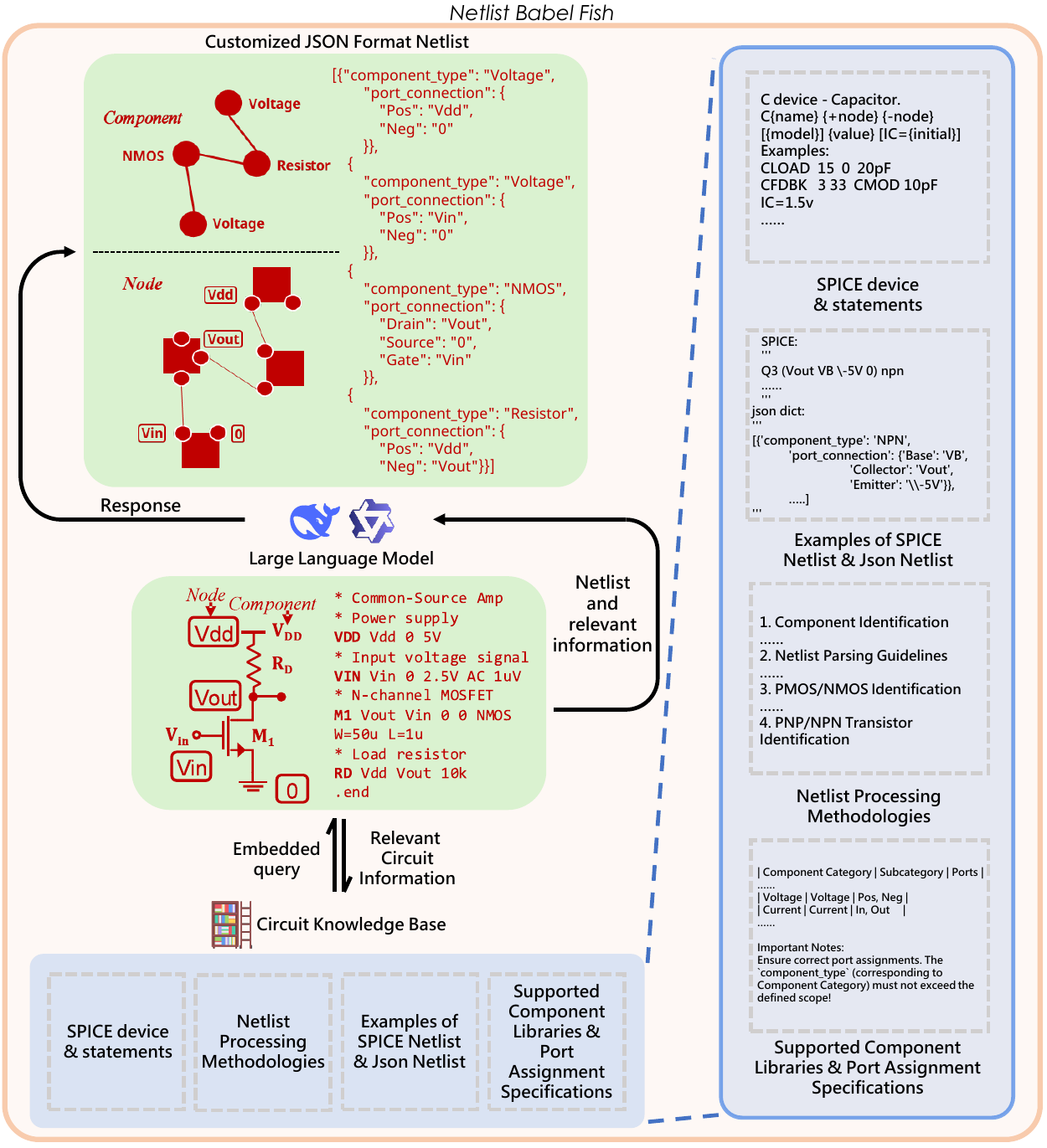}

\caption{Workflow of Netlist Babel Fish. First, an embedded query system processes the SPICE netlist by retrieving relevant information from our circuit knowledge base, which covers netlist preprocessing methodologies, supported component libraries including port assignment specifications, and exemplar netlists in both SPICE and JSON formats. Then, an LLM receives the netlist along with retrieved data and generates the corresponding JSON netlist as output.}\label{NetlistBabelFish}

\end{figure*}

To address the challenge of differing dataset formats, we develop Netlist Babel Fish, a framework that enables bidirectional conversion between netlist formats by integrating domain-specific information. Netlist Babel Fish integrates an LLM with a RAG system, specifically LangChain, grounded in a structured SPICE knowledge base. This knowledge base includes brief references to the SPICE devices and statements \cite{ecircuitcenterSPICECommands}, methodologies for the netlist process, supported component libraries with port assignment specifications, and examples illustrating practical implementations. We individually evaluate DeepSeek-V2.5 \cite{deepseekai2024deepseekv2strongeconomicalefficient, deepseekDeepSeekV25OpenSource}, DeepSeek-V3-0324 \cite{deepseekai2025deepseekv3technicalreport, deepseekDeepSeekV30324}, DeepSeek-V3.1 \cite{deepseekDeepSeekV31}, and Qwen3-30B-A3B \cite{yang2025qwen3technicalreport} as the LLM in this framework, all of which proved competent. We show the workflow of Netlist Babel Fish converting a SPICE netlist to a custom JSON format \cite{eda2024} in Figure \ref{NetlistBabelFish}. This workflow is also reversible for conversion from the custom JSON format to SPICE.

%%% newest version v19

The current implementation focuses on converting between SPICE and JSON formats. However, its modular architecture theoretically supports arbitrary netlist format conversions with appropriate domain knowledge bases. The system's extensibility stems from its decoupled knowledge representation framework, where format-specific conversion rules are explicitly encoded. We demonstrate converting SPICE netlists into Verilog-A and VHDL-AMS formats in Appendix \ref{extensibility_NBF}.

\section{SpiceNetlist: Datasets Rectification}
\label{SpiceNetlist+Preprocessing}
Regarding the challenge of the circuit link prediction problem, there are few open datasets available. Moreover, existing SPICE netlist datasets often contain content errors and syntactic inconsistencies, which lead to incompatibility with Netlist Babel Fish for proper parsing.

\begin{figure*}%% placement specifier
\centering%% For centre alignment of image.
\includegraphics[width=0.95\linewidth]{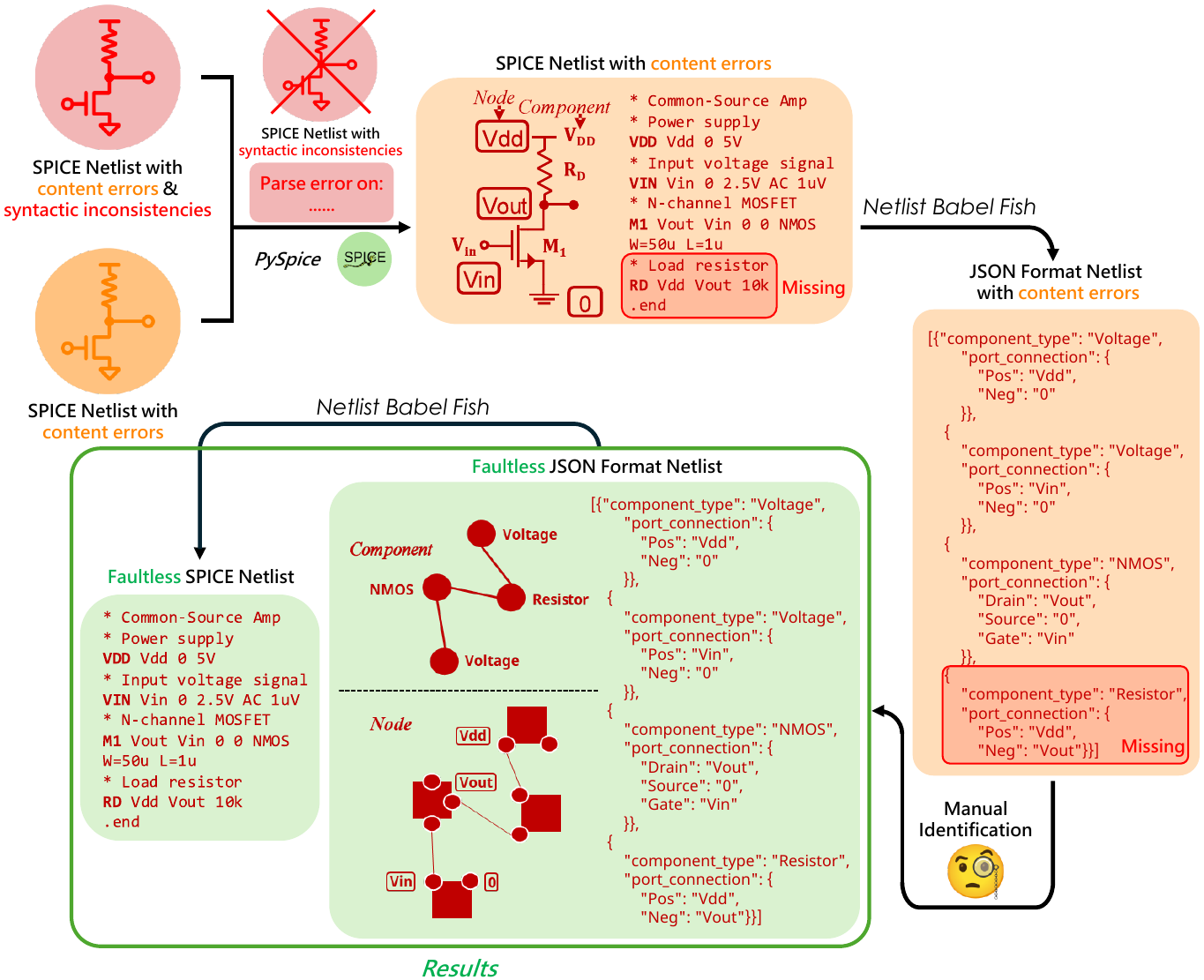}

\caption{SPICE netlist datasets preprocessing pipeline. First, we utilize PySpice to validate SPICE netlists and filter out any that are syntactically inconsistent. Then, the remaining netlists are converted to JSON format using Netlist Babel Fish, where content errors are manually corrected. The resulting faultless JSON netlists are then processed back through Netlist Babel Fish to generate faultless SPICE equivalents.}

\label{preprocessing_pipeline}
\end{figure*}

To address this issue, we utilize PySpice \cite{PySpice} to automatically detect and filter out SPICE netlists with syntactic inconsistencies. Subsequently, we convert all datasets to JSON format with Netlist Babel Fish and perform manual verification to identify and rectify netlists with content errors in both formats. We demonstrate this preprocessing pipeline~in~Figure~\ref{preprocessing_pipeline}. 

Following this pipeline, we integrate two datasets, Kicad Github \cite{said_circuit_2023} and AMSNet \cite{tao2024amsnetnetlistdatasetams} as one unified large-scale dataset, SpiceNetlist. We present its statistics in Table \ref{spicenetlist}-\ref{type_number_spicenetlist}. Furthermore, we apply the preprocessing pipeline to three additional datasets: Image2Net\footnote{Image2Net is provided by \href{mailto:panyiren@hdu.edu.cn}{Yiren Pan (panyiren@hdu.edu.cn)}, who is currently writing a paper about his findings, and originated from the 2024 China Postgraduate IC Innovation Competition - EDA Elite Challenge Contest \cite{edadataset} Image Dataset.}, Masala-CHAI \cite{bhandari2025masalachailargescalespicenetlist} and AnalogGenie \cite{gao2025analoggenie}.

\section{Experiments}
%% Labels are used to cross-reference an item using \ref command.

\subsection{Dataset Processing}
\label{preprocessing}
In our experiments, we utilize SpiceNetlist, Image2Net, Masala-CHAI, and AnalogGenie\footnote{Given the substantial size of AnalogGenie, we randomly select a representative 10\% subset for our experiments.}. Table \ref{Image2Net} shows the statistics of all datasets, and Table \ref{comp} lists all components featured by all datasets.

% \begin{table}
% \centering
% \caption{Statistics of all datasets utilized in our experiments.}
% \resizebox{0.55\textwidth}{!}{
% \begin{tabular}{lccccc}
% \hline
% \textbf{Dataset} & \textbf{Graphs} & \textbf{Classes} & \textbf{Avg. Nodes} & \textbf{Avg. Edges} \\
% \hline
% SpiceNetlist & 775 & 10 & 13.78 & 15.60 \\
% Image2Net & 1960 & 13 & 26.36 & 37.80 \\
% Masala-CHAI & 4057 & 13 & 15.72 & 19.97 \\
% AnalogGenie & 4064 & 9 & 63.89 & 168.73 \\
% \hline
% \end{tabular}
% }
% \label{Image2Net}
% \end{table}

\begin{table}
\centering
\begin{minipage}{\linewidth}
  % ---------- 左表 ----------
  \begin{minipage}[t]{0.42\linewidth}
    \vspace{0pt}% 锚点：让表格贴顶
    \centering
    \caption{Types and number of circuits in SpiceNetlist.}
    \resizebox{\linewidth}{!}{%
      \begin{tabular}{lc}
        \hline
        \textbf{Type} & \textbf{Number} \\
        \hline
        Current Mirrors and Sources        & 387 \\
        Amplifiers                         & 258 \\
        Other                              &  51 \\
        Filters and Passive Networks       &  41 \\
        Characterization and Test Circuits &  19 \\
        Oscillators                        &  18 \\
        Voltage and Power Management       &  15 \\
        \hline
      \end{tabular}%
    }
    \label{type_number_spicenetlist}
  \end{minipage}\hfill
  % ---------- 右表 ----------
  \begin{minipage}[t]{0.56\linewidth}
  \vspace{0pt}% 锚点：让表格贴顶
    \centering
    \caption{Statistics of all datasets utilized in our experiments.}
    \resizebox{\linewidth}{!}{%
      \begin{tabular}{lccccc}
        \hline
        \textbf{Dataset} & \textbf{Graphs} & \textbf{Classes} & \textbf{Avg.\ Nodes} & \textbf{Avg.\ Edges} \\
        \hline
        SpiceNetlist &  775 & 10 & 13.78 &  15.60 \\
        Image2Net    & 1960 & 13 & 26.36 &  37.80 \\
        Masala-CHAI  & 4057 & 13 & 15.72 &  19.97 \\
        AnalogGenie  & 4064 &  9 & 63.89 & 168.73 \\
        \hline
      \end{tabular}%
    }
    \label{Image2Net}
  \end{minipage}
\end{minipage}
\end{table}

\begin{table*}
\footnotesize
\centering
\caption{Component types and their labels/ports in SpiceNetlist, Image2Net, Masala-CHAI and AnalogGenie.}
% \resizebox{0.8\textwidth}{!}{%
% \begin{tabular}{lll}
\begin{tabular*}{\textwidth}{@{\extracolsep{\fill}}lll}
\hline
\textbf{Component} & \textbf{Label} & \textbf{Ports} \\
\hline
PMOS/NMOS & PMOS/NMOS & Drain, Source, Gate \\
Voltage Source & Voltage & Pos, Neg \\
Current Source & Current & In, Out \\
BJT (NPN/PNP) & NPN/NPN\_cross/PNP/PNP\_cross & Base, Emitter, Collector \\
Diode & Diode & In, Out \\
DISO Amplifier & Diso\_amp & InN, InP, Out \\
SISO Amplifier & Siso\_amp & In, Out \\
DIDO Amplifier & Dido\_amp & InN, InP, OutN, OutP \\
Passive Components (Cap, Ind, Res) & Cap, Ind, Res & Pos, Neg \\
\hline
% \end{tabular}}
\end{tabular*}
% }
\label{comp}
\end{table*}

Following practices similar to batching in the graph learning domain, we represent the entire dataset as a comprehensive graph by combining multiple adjacency matrices through block-diagonal concatenation. Define ${{A}}_i \in [0,1]^{N_i \times N_i}$ as the adjacency matrix of $G_i$, where $N_i$ denoting vertex count. Then, the combined graph's total vertex count $N$ becomes $\sum _{1\le j\le n } N_j$. The stacked adjacency matrix $A \in \mathbb{R}^{N \times N}$ and concatenated node characteristic vector are defined as:

\begin{equation} % only appear once
{A} = \begin{bmatrix}
{A}_{1} & & \\
& \ddots & \\
& & {A}_{n}
\end{bmatrix}, \quad
{X} = \begin{bmatrix}
{X}_1 \\
\vdots \\
{X}_n
\end{bmatrix}
\label{eq:stacked_adjacency_matrix_vector}
\end{equation}

State-of-the-art approaches for link prediction commonly expect a unified graph representation to learn the network structure \cite{veličković2018graphattentionnetworks, jiang2020coembeddingnodesedgesgraph}. Combining all training graphs allows seamless integration with established GNN methods while maintaining computational efficiency. The comprehensive graph also efficiently leverages sparse data structures \cite{2022150, qarkaxhija2024link, liang2025can}. \label{stack}

\subsection{Experimental Setup}

We conduct two distinct evaluation paradigms: intra-dataset evaluation and cross-dataset evaluation. For intra-dataset evaluation, we utilize the identical dataset for both model training and testing, specifically SpiceNetlist, Image2Net, and Masala-CHAI \cite{bhandari2025masalachailargescalespicenetlist}. For cross-dataset evaluation, aiming to assess the generalizability of our approach, we conduct training on a single dataset and testing on another dataset from the three, with additional verification using AnalogGenie.

For both evaluation paradigms, we employ two experimental configurations: one using the conventional dataset splitting strategy (train, validation, and test splits) and another employing the 5-fold cross-validation strategy, given the prevalent challenge of limited dataset sizes. Here, the traditional split serves as an ablation study to assess the impact of 5-fold cross-validation, ensuring robustness in performance evaluation. For both configurations, we maintain a fixed data split ratio of 70\% training, 20\% test, and 10\% validation. We also conduct experiments using \cite{said_circuit_2023} as the baseline, which employs the conventional dataset splitting strategy.

We adopt the SEAL framework with the following PyTorch Geometric implementation: batch size = $1$, $2$-hop neighborhoods, and a maximum of $50$ training epochs with early stopping. The early stopping criterion only engages once test accuracy shows improvement and exceeds $0.5000$. Thereafter, we terminate the training if we observe no improvement of $\geq0.0001$ for three consecutive epochs. We use the one-hot encoding of node labels generated through DRNL and concatenate them with the one-hot encoding of the component types. For SpiceNetlist, we set the learning rate to $1e-6$. For Image2Net and Masala-CHAI, we set the learning rate to $1e-6$ for the baseline and $6e-8$ for~our~method.

% We construct the training graph using the adjacency stacking approach from Section \ref{stack}. During each experiment, we randomly select a vertex from every test graph and predict its connections. After 10 repetitions, we calculate and report the average accuracy, ROC-AUC and inference time\footnote{Inference time includes the time for subgraph extraction, encoding and GNN inference on all subgraphs of the dataset's test split.}.
We construct the training graph using the adjacency stacking approach from Section \ref{stack}. During each experiment, we randomly select a vertex from every test graph and predict its connections. After 10 repetitions, we calculate and report the average accuracy and ROC-AUC.

\subsection{Results and Analysis} 

The experimental results demonstrate substantial outperformance of our method across all evaluation scenarios:

In the intra-dataset evaluation, \textbf{1)} We achieve accuracy gains of 11.08\% (conventional split) and 16.08\% (5-fold cross-validation) on SpiceNetlist; \textbf{2)} We achieve accuracy gains of 9.18\% (conventional split) and 11.38\% (5-fold cross-validation) on Image2Net; \textbf{3)} We achieve accuracy gains of 13.89\% (conventional split) and 16.02\% (5-fold cross-validation) on Masala-CHAI. We present the results in Table \ref{Comparison} and Figure \ref{tendency_intra_dataset_experiment}. 

In the cross-dataset evaluation, our work maintains accuracy ranging from 92.05\% (trained on Image2Net, tested on SpiceNetlist) to 99.07\% (trained on SpiceNetlist, tested on AnalogGenie), demonstrating its strong capability to transfer learned features across different circuit netlist datasets. This generalizability stems from its design, which prioritizes capturing invariant structural and semantic features in analog circuit graphs, enabling the model to generalize effectively, even with variations in dataset scales and complexities. Notably, our design aligns with Google DeepMind's AlphaChip \cite{mirhoseini2021graph}. For each edge, SEAL extracts a k-hop enclosing sub-graph and labels each node using DRNL, which captures local topological motifs. Therefore, GNN-ACLP focuses on relationships rather than relying on absolute net names or global coordinates. Additionally, SEAL's embeddings incorporate native characteristics of component nodes, which are concatenated into the input representations, allowing the model to learn rich, low-dimensional embeddings that encode both structural connectivity and semantic roles. These embeddings implicitly distill universal principles of analog design, such as hierarchical port connections and device functionalities, making them resilient to differences across dataset sources. As a result, SEAL propagates information iteratively and fosters representations that generalize well. We present the results in Table \ref{cross-verification}. 

Our work yields promising results, indicating that the proposed method effectively learns netlist structures and can be applied to component link prediction tasks.

\begin{table*}
\centering
\caption{Accuracy and ROC-AUC in intra-dataset evaluation. The bold \textbf{\textcolor{red}{red}} and \textbf{\textcolor{blue}{blue}} fonts denote the best and the sub-optimal metrics, respectively.}
\resizebox{\textwidth}{!}{
\begin{tabular}{@{\extracolsep{\fill}}l c c c c c c}
\hline
\multirow{2}{*}{Method} & \multicolumn{2}{c}{SpiceNetlist} & \multicolumn{2}{c}{Image2Net} & \multicolumn{2}{c}{Masala-CHAI} \\
\cline{2-7}
 & Accuracy & ROC-AUC & Accuracy & ROC-AUC & Accuracy & ROC-AUC \\
\hline
Baseline & 77.96\% ± 1.71\% & 0.8774 ± 0.0056 & 85.72\% ± 1.95\% & 0.9367 ± 0.0164 & 80.40\% ± 0.92\% & 0.8995 ± 0.0099 \\
\textbf{Conv. Split (Ours)} & \textbf{\textcolor{blue}{89.04\% ± 1.18\%}} & \textbf{\textcolor{blue}{0.9645 ± 0.0097}} & \textbf{\textcolor{blue}{94.90\% ± 0.25\%}} & \textbf{\textcolor{blue}{0.9926 ± 0.0004}} & \textbf{\textcolor{blue}{94.29\% ± 0.28\%}} & \textbf{\textcolor{blue}{0.9892 ± 0.0009}} \\
\textbf{5-fold CV (Ours)} & \textbf{\textcolor{red}{94.04\% ± 0.70\%}} & \textbf{\textcolor{red}{0.9866 ± 0.0019}} & \textbf{\textcolor{red}{97.10\% ± 0.24\%}} & \textbf{\textcolor{red}{0.9979 ± 0.0003}} & \textbf{\textcolor{red}{96.42\% ± 1.24\%}} & \textbf{\textcolor{red}{0.9959 ± 0.0005}} \\
\hline
\end{tabular}}
\label{Comparison}
\end{table*}

\begin{figure}
  \centering
  \begin{minipage}{0.96\textwidth}
    \begin{subfigure}{\textwidth}
      \includegraphics[width=\linewidth]{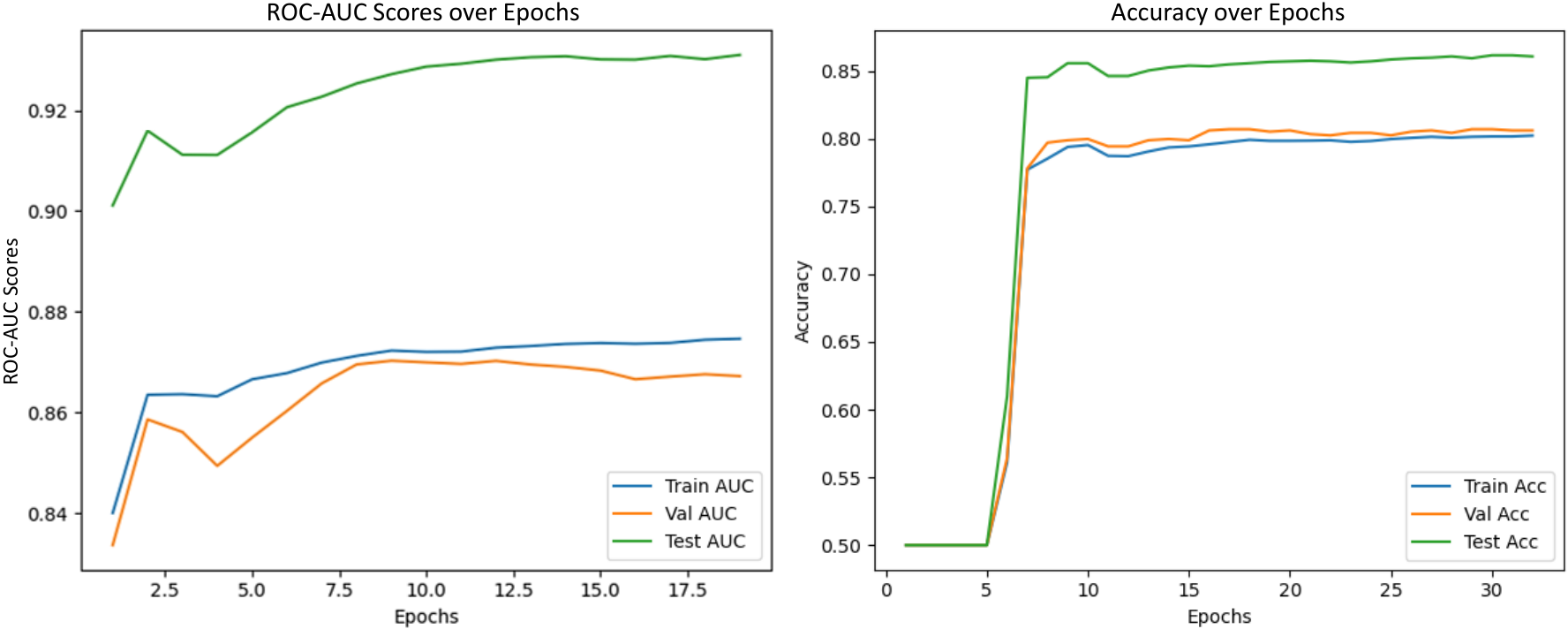}
      \subcaption{ROC-AUC and accuracy variation tendencies observed in the baseline. Both indicators vibrate violently.}
      \label{original_acc}
    \end{subfigure}

    \vspace{1.0em}
    \begin{subfigure}{\textwidth}
      \includegraphics[width=\linewidth]{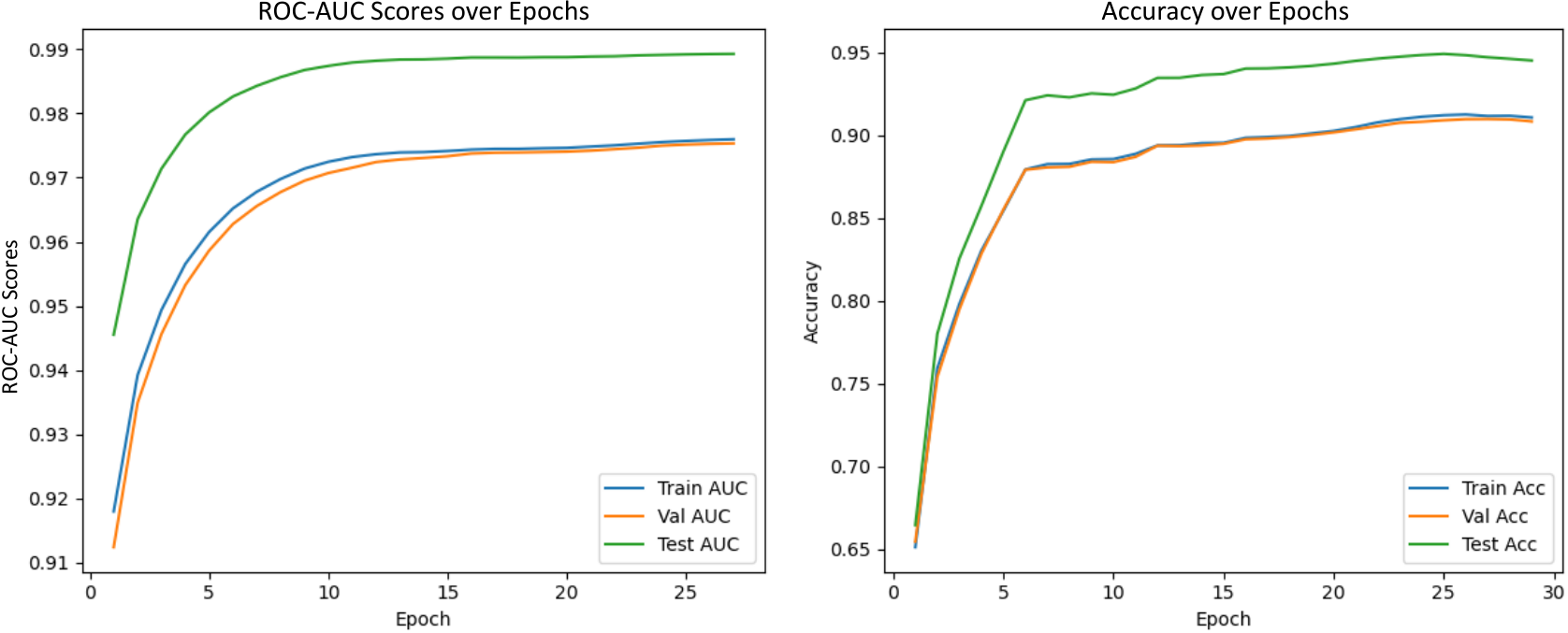}
      \subcaption{ROC-AUC and accuracy variation tendencies observed in our method with the conventional dataset splitting strategy. Both indicators vibrate less violently than the baseline method.}
      \label{new_acc}
    \end{subfigure}
  \end{minipage}
  \caption{ROC-AUC and accuracy variation tendencies observed in the intra-dataset experiment.}
  \label{tendency_intra_dataset_experiment}
\end{figure}

\begin{figure}
  \ContinuedFloat
  \centering
  \begin{minipage}{0.96\textwidth}
    \begin{subfigure}{\textwidth}
      \includegraphics[width=\linewidth]{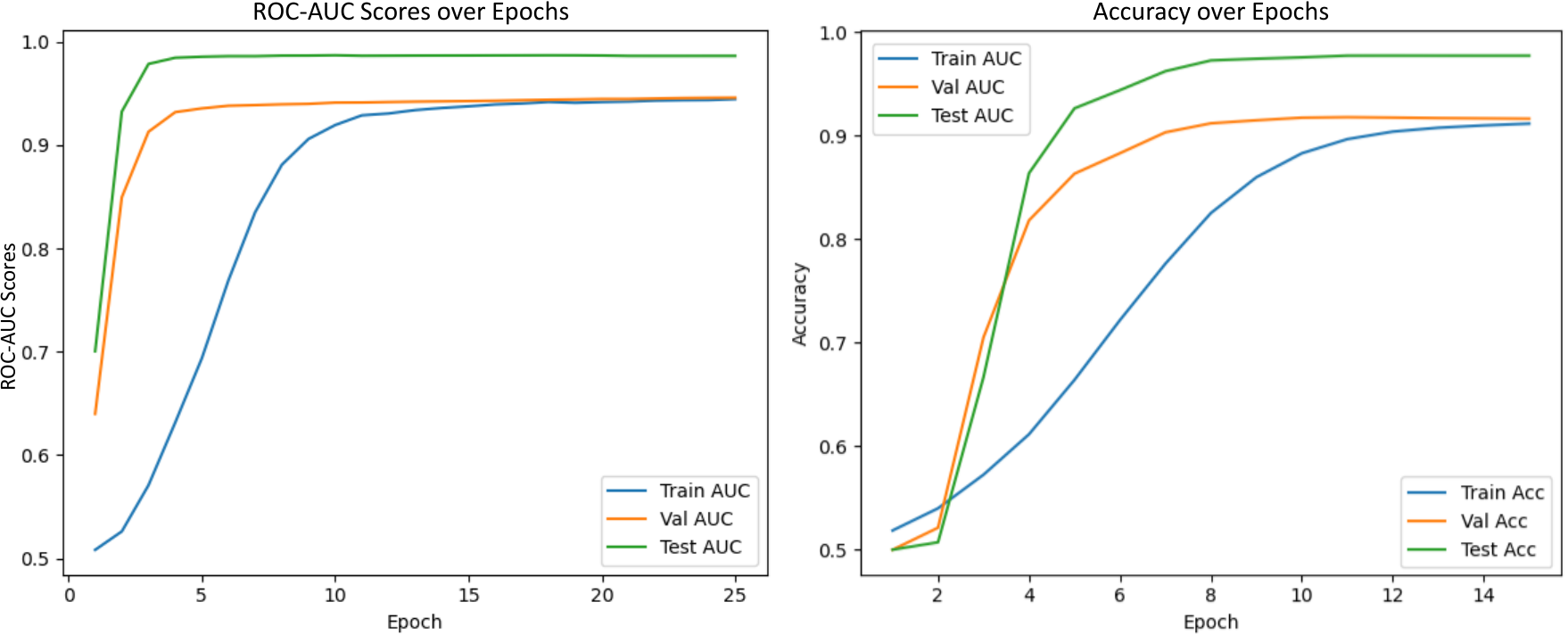}
      \subcaption{ROC-AUC and accuracy variation tendencies observed in our method with the 5-fold cross-validation strategy. Both indicators increase steadily.}
      \label{kfold_acc}
    \end{subfigure}
  \end{minipage}
  \caption[]{ROC-AUC and accuracy variation tendencies observed in the intra-dataset experiment (cont.)}
  \label{tendency_intra_dataset_experiment_cont}
\end{figure}

% cross-verification

\begin{table*}
\tiny
  \centering
  \caption{Accuracy and ROC-AUC in cross-dataset evaluation. The \textbf{bold} fonts denote the best results.}
  \label{cross-verification}
  \resizebox{1\linewidth}{!}{%
  \begin{tabular}{c@{\hspace{0.3em}}ccccc} % 6 columns
    \toprule
    \multirow{2}{*}{Train Dataset} & \multirow{2}{*}{Test Dataset} & \multicolumn{2}{c}{Baseline Performance} & \multicolumn{2}{c}{\textbf{5-fold CV (Ours) Performance}} \\
    \cmidrule(lr){3-4} \cmidrule(lr){5-6}
    & & Accuracy & ROC-AUC & Accuracy & ROC-AUC \\
    \midrule
    SpiceNetlist & Image2Net & 82.95\% ± 1.45\% & 0.9163 ± 0.0114 & \textbf{96.94\% ± 0.88\%} & \textbf{0.9963 ± 0.0014} \\
    SpiceNetlist & Masala-CHAI & 77.00\% ± 1.47\% & 0.8600 ± 0.0030 & \textbf{95.98\% ± 1.08\%} & \textbf{0.9945 ± 0.0017} \\
    Image2Net & SpiceNetlist & 75.10\% ± 3.11\% & 0.8848 ± 0.0050 & \textbf{92.05\% ± 0.75\%} & \textbf{0.9867 ± 0.0021} \\
    Image2Net & Masala-CHAI & 75.47\% ± 1.28\% & 0.8706 ± 0.0048 & \textbf{96.12\% ± 0.66\%} & \textbf{0.9962 ± 0.0007} \\
    Masala-CHAI & SpiceNetlist & 72.84\% ± 8.50\% & 0.8678 ± 0.0424 & \textbf{93.06\% ± 1.66\%} & \textbf{0.9859 ± 0.0019} \\
    Masala-CHAI & Image2Net & 82.35\% ± 8.01\% & 0.9293 ± 0.0151 & \textbf{97.18\% ± 0.85\%} & \textbf{0.9974 ± 0.0004} \\
    SpiceNetlist & AnalogGenie & 69.99\% ± 2.17\% & 0.7822 ± 0.0330 & \textbf{99.07\% ± 0.82\%} & \textbf{0.9995 ± 0.0003} \\
    Image2Net & AnalogGenie & 71.23\% ± 2.32\% & 0.7950 ± 0.0402 & \textbf{98.73\% ± 1.32\%} & \textbf{0.9997 ± 0.0002} \\
    Masala-CHAI & AnalogGenie & 74.19\% ± 2.58\% & 0.8107 ± 0.0378 & \textbf{98.68\% ± 1.07\%} & \textbf{0.9997 ± 0.0002} \\
    \bottomrule
  \end{tabular}%
  } % Closing brace for \resizebox
\end{table*}

\subsection{Ablation Study} 

To demonstrate the effectiveness of the SEAL framework underlying our method, we perform an ablation study by replacing the SEAL framework with several alternative methods, including heuristic methods, latent feature methods, and other GNN-based methods, across SpiceNetlist, Image2Net, and Masala-CHAI, using the 5-fold cross-validation strategy. We present the results in Table \ref{ablation}\footnote{\label{OOMOOT}Since some of the frameworks are infeasible due to high computational complexity, we present results only for implementable methods (i.e., OOM: >48GB VRAM required; OOT: >48 hour training time).}.

\begin{table*}
     \caption{Ablation study comparison results of SEAL and other methods. The bold \textbf{\textcolor{red}{red}} and \textbf{\textcolor{blue}{blue}} fonts denote the best and the sub-optimal results, respectively.}
    \begin{subtable}{1\textwidth}
        \centering
        \resizebox{\linewidth}{!}{%
        \begin{tabular}{lccccccccc}
        \hline
        Framework & \multicolumn{2}{c}{SpiceNetlist} & \multicolumn{2}{c}{Image2Net} & \multicolumn{2}{c}{Masala-CHAI} \\
        \cline{2-7}
        & Accuracy & ROC-AUC & Accuracy & ROC-AUC & Accuracy & ROC-AUC \\
        \hline
        common neighbors \cite{PhysRevE.64.025102_CN_preferential_attachment} & 75.86\% ± 0.40\% & 0.7586 ± 0.0040 & 88.36\% ± 0.15\% & 0.8836 ± 0.0015 & 86.16\% ± 0.16\% & 0.8616 ± 0.0016 \\
        Jaccard \cite{pn2005introduction_Jaccard} & 74.92\% ± 0.89\% & 0.7586 ± 0.0040 & 88.34\% ± 0.16\% & 0.8836 ± 0.0015 & 86.14\% ± 0.15\% & 0.8616 ± 0.0016 \\
        preferential attachment \cite{barabasi1999emergence_preferential_attachment} & 59.16\% ± 1.74\% & 0.6198 ± 0.0059 & 65.33\% ± 0.42\% & 0.6972 ± 0.0023 & 65.65\% ± 0.11\% & 0.7043 ± 0.0023 \\
        Adamic-Adar \cite{adamic2003friends_aa} & 75.82\% ± 0.41\% & 0.7586 ± 0.0040 & 88.35\% ± 0.15\% & 0.8836 ± 0.0015 & 86.16\% ± 0.15\% & 0.8616 ± 0.0016 \\
        resource allocation \cite{zhou2009predicting_aa_ra} & 75.82\% ± 0.41\% & 0.7586 ± 0.0040 & 88.35\% ± 0.15\% & 0.8836 ± 0.0015 & 86.16\% ± 0.15\% & 0.8616 ± 0.0016 \\
        Katz \cite{katz1953new_katz} & \textbf{\textcolor{blue}{85.08\% ± 0.32\%}} & \textbf{\textcolor{blue}{0.8515 ± 0.0033}} & OOT & OOT & OOT & OOT \\
        PageRank \cite{brin1998anatomy_pagerank} & 59.79\% ± 1.47\% & 0.6277 ± 0.0059 & OOT & OOT & OOT & OOT \\
        SimRank \cite{jeh2002simrank_simrank} & 79.47\% ± 0.16\% & 0.7952 ± 0.0015 & \textbf{\textcolor{blue}{90.95\% ± 0.03\%}} & 0.9096 ± 0.0004 & OOM & OOM \\
        WLNM \cite{zhang2017weisfeiler_WLNM} & 69.33\% ± 0.36\% & 0.8170 ± 0.0025 & 89.15\% ± 0.03\% & \textbf{\textcolor{blue}{0.9779 ± 0.0002}} & \textbf{\textcolor{blue}{91.22\% ± 0.04\%}} & \textbf{\textcolor{blue}{0.9757 ± 0.0001}} \\
        WEGL \cite{kolouri2021wasserstein_WEGL} & 64.17\% ± 3.26\% & 0.6036 ± 0.0384 & 80.92\% ± 2.46\% & 0.8230 ± 0.0207 & 77.53\% ± 4.71\% & 0.7752 ± 0.0391 \\
        \textbf{SEAL (Ours)} & \textbf{\textcolor{red}{94.04\% ± 0.70\%}} & \textbf{\textcolor{red}{0.9866 ± 0.0019}} & \textbf{\textcolor{red}{97.10\% ± 0.24\%}} & \textbf{\textcolor{red}{0.9979 ± 0.0003}} & \textbf{\textcolor{red}{96.42\% ± 1.24\%}} & \textbf{\textcolor{red}{0.9959 ± 0.0005}} \\
        \hline
        \end{tabular}
        }
        \caption{Ablation study comparison results of SEAL and heuristic methods.}
        % \label{ablation_heuristic}
    \end{subtable}
    \par\vspace{0.6em}
    \begin{subtable}{\textwidth}
        \centering
        \resizebox{\linewidth}{!}{%
        \begin{tabular}{lccccccccc}
        \hline
        Framework & \multicolumn{2}{c}{SpiceNetlist} & \multicolumn{2}{c}{Image2Net} & \multicolumn{2}{c}{Masala-CHAI} \\
        \cline{2-7}
        & Accuracy & ROC-AUC & Accuracy & ROC-AUC & Accuracy & ROC-AUC \\
        \hline
        SBM \cite{aicher2015learning_SBM} & 58.57\% ± 1.75\% & 0.6109 ± 0.0031 & 59.77\% ± 0.11\% & 0.6032 ± 0.0015 & 60.91\% ± 0.07\% & 0.6144 ± 0.0013 \\
        BMF \cite{bhavana2024matrix_2024} & 52.26\% ± 0.19\% & 0.5405 ± 0.0031 & 58.96\% ± 0.09\% & 0.6760 ± 0.0008 & 58.23\% ± 0.08\% & 0.6621 ± 0.0010 \\
        graph2vec \cite{mlg2017_21_graph2vec} & 65.75\% ± 2.74\% & 0.7315 ± 0.0283 & 60.69\% ± 0.41\% & 0.6938 ± 0.0089 & OOT & OOT \\
        AROPE \cite{10.1145/3219819.3219969_AROPE} & \textbf{\textcolor{blue}{71.56\% ± 0.15\%}} & 0.8048 ± 0.0027 & \textbf{\textcolor{blue}{71.50\% ± 0.08\%}} & \textbf{\textcolor{blue}{0.8041 ± 0.0010}} & \textbf{\textcolor{blue}{71.52\% ± 0.14\%}} & \textbf{\textcolor{blue}{0.8044 ± 0.0022}} \\
        SPC \cite{ng2001spectral_SPC} & 71.53\% ± 3.26\% & \textbf{\textcolor{blue}{0.8197 ± 0.0108}} & OOT & OOT & OOT & OOT \\
        \textbf{SEAL (Ours)} & \textbf{\textcolor{red}{94.04\% ± 0.70\%}} & \textbf{\textcolor{red}{0.9866 ± 0.0019}} & \textbf{\textcolor{red}{97.10\% ± 0.24\%}} & \textbf{\textcolor{red}{0.9979 ± 0.0003}} & \textbf{\textcolor{red}{96.42\% ± 1.24\%}} & \textbf{\textcolor{red}{0.9959 ± 0.0005}} \\
        \hline
        \end{tabular}
        }
        \caption{Ablation study comparison results of SEAL and latent feature methods.}
        % \label{ablation_latent}
    \end{subtable}
    \par\vspace{0.6em}
    \begin{subtable}{\textwidth}
        \centering
        \resizebox{\linewidth}{!}{%
        \begin{tabular}{lccccccccc}
        \hline
        Framework & \multicolumn{2}{c}{SpiceNetlist} & \multicolumn{2}{c}{Image2Net} & \multicolumn{2}{c}{Masala-CHAI} \\
        \cline{2-7}
        & Accuracy & ROC-AUC & Accuracy & ROC-AUC & Accuracy & ROC-AUC \\
        \hline
        VGAE \cite{kipf2016variational_VGAE} & 63.80\% ± 2.00\% & 0.7197 ± 0.0099 & 71.58\% ± 1.18\% & 0.8318 ± 0.0078 & 69.37\% ± 1.54\% & 0.8115 ± 0.0114 \\
        GraphSAGE \cite{hamilton2017inductive} & \textbf{\textcolor{blue}{65.07\% ± 2.94\%}} & \textbf{\textcolor{blue}{0.7683 ± 0.0172}} & \textbf{\textcolor{blue}{78.45\% ± 0.26\%}} & \textbf{\textcolor{blue}{0.9147 ± 0.0025}} & \textbf{\textcolor{blue}{77.74\% ± 0.19\%}} & \textbf{\textcolor{blue}{0.9108 ± 0.0009}} \\
        GCN+LRGA \cite{puny2020global} & 50.89\% ± 1.21\% & 0.7029 ± 0.0268 & OOT & OOT & OOT & OOT \\
        \textbf{SEAL (Ours)} & \textbf{\textcolor{red}{94.04\% ± 0.70\%}} & \textbf{\textcolor{red}{0.9866 ± 0.0019}} & \textbf{\textcolor{red}{97.10\% ± 0.24\%}} & \textbf{\textcolor{red}{0.9979 ± 0.0003}} & \textbf{\textcolor{red}{96.42\% ± 1.24\%}} & \textbf{\textcolor{red}{0.9959 ± 0.0005}} \\
        \hline
        \end{tabular}
        }
        \caption{Ablation study comparison results of SEAL and other GNN-based methods.}
        % \label{ablation_gnn}
     \end{subtable}
     \label{ablation}
\end{table*}

We attribute the superior performance of SEAL to its inherent adaptability for predicting links in analog circuits, where the underlying graphs show both sparsity and large-scale topology.

Heuristic methods, which rely on closed-triangle assumptions or preferential attachment mechanisms, prove inadequate in sparse graphs due to their inability to capture structural idiosyncrasies \cite{shang2019link}. Latent feature methods fail either since insufficient node interactions prevent meaningful learning of high-order correlations needed for latent semantic extraction \cite{lu2011latent}. They may also erroneously encode unobserved links as zero-weight edges, creating inconsistency between partial and complete graph representations \cite{yokoi2017link}. As for other GNN-based methods here, they are node-based and compute two node representations independently of each other without considering their relative positions and associations. SEAL, on the other hand, is a subgraph-based method and extracts an enclosing subgraph around each target link. Therefore, it better models the associations between two target nodes in a sparse graph. The node labeling step in SEAL also helps model the associations between the two target nodes, especially when the graph~is~sparse~\cite{Zhang2022}.

The computational resources required by different frameworks for predicting on large circuit graphs are also a critical consideration. Katz \cite{zhu2025finding}, PageRank \cite{wang2017ranking}, SimRank \cite{ge2023efficient}, graph2vec \cite{szakacswhole}, SPC \cite{he2021scalable}, GCN \cite{blakely2021time} and VGAE\footnote{VGAE faces a trade-off in node feature decomposition: a large latent dimension increases computational overhead, while a small one may fail to capture essential latent factors \cite{fu2023variational}. Additionally, VGAE-based methods often suffer from training instability, and correcting this deviation introduces additional computational costs. \cite{e22020197, chen2024preserving}} suffer from high computational complexity, making them inefficient for large-scale circuit graphs. Therefore, when applied to large circuit graphs, these methods exhibit prohibitive resource demands and may even OOM or OOT. For instance, SimRank consumes nearly 40 GB of VRAM on Image2Net, graph2vec requires more than 20 hours for training on SpiceNetlist, and VGAE occupies 34 GB of RAM on Masala-CHAI. In contrast, SEAL completes training on each dataset (SpiceNetlist, Image2Net, and Masala-CHAI) in under 10 hours utilizing significantly less VRAM.

\section{Discussion}

While our GNN-ACLP showcases significant promise in circuit link prediction, limitations exist. To comprehensively analyze the limits of GNN-ACLP, failed examples are shown in Table \ref{discussion}\footnotemark[\getrefnumber{OOMOOT}]. Specifically, we train and test GNN-ACLP and other methods on AnalogGenie with the 5-fold cross-validation strategy.

\begin{table}
    \caption{Performance comparison results on AnalogGenie. The bold \textbf{\textcolor{red}{red}} and \textbf{\textcolor{blue}{blue}} fonts denote the best and the sub-optimal results,~respectively.}
    \label{discussion}
    \begin{minipage}[t]{0.53\textwidth}
    \vspace{0pt}
        \begin{subtable}{\linewidth}
            \centering
            \resizebox{\linewidth}{!}{%
            \begin{tabular}{lcc}
            \hline
            Framework & \multicolumn{2}{c}{AnalogGenie} \\
            \cline{2-3}
            & Accuracy & ROC-AUC \\
            \hline
            common neighbors & \textbf{\textcolor{blue}{95.32\% ± 0.03\%}} & \textbf{\textcolor{blue}{0.9532 ± 0.0003}} \\
            Jaccard & \textbf{\textcolor{blue}{95.32\% ± 0.03\%}} & \textbf{\textcolor{blue}{0.9532 ± 0.0003}} \\
            preferential attachment & 69.64\% ± 0.05\% & 0.7403 ± 0.0006 \\
            Adamic-Adar & \textbf{\textcolor{blue}{95.32\% ± 0.03\%}} & \textbf{\textcolor{blue}{0.9532 ± 0.0003}} \\
            resource allocation & \textbf{\textcolor{blue}{95.32\% ± 0.03\%}} & \textbf{\textcolor{blue}{0.9532 ± 0.0003}} \\
            Katz & OOT & OOT \\
            PageRank & OOT & OOT \\
            SimRank & OOM & OOM \\
            WLNM & \textbf{\textcolor{red}{97.46\% ± 0.04\%}} & \textbf{\textcolor{red}{0.9978 ± 0.0000}} \\
            WEGL & 82.69\% ± 3.14\% & 0.8778 ± 0.0295 \\
            \textbf{SEAL (Ours)} & OOT & OOT \\
            \hline
            \end{tabular}
            }
            \caption{Performance comparison results of SEAL and heuristic methods on AnalogGenie.}
            \label{subtable:heuristic}
        \end{subtable}
    \end{minipage}\hfill
    \begin{minipage}[t]{0.45\textwidth}
    \vspace{0pt}
        \begin{subtable}{\linewidth}
            \centering
            \resizebox{\linewidth}{!}{%
            \begin{tabular}{lcc}
            \hline
            Framework & \multicolumn{2}{c}{AnalogGenie} \\
            \cline{2-3}
            & Accuracy & ROC-AUC \\
            \hline
            SBM & 56.05\% ± 0.02\% & 0.5629 ± 0.0004 \\
            BMF & \textbf{\textcolor{blue}{64.38\% ± 0.03\%}} & \textbf{\textcolor{blue}{0.7776 ± 0.0002}} \\
            graph2vec & OOT & OOT \\
            AROPE & \textbf{\textcolor{red}{71.55\% ± 0.20\%}} & \textbf{\textcolor{red}{0.8046 ± 0.0027}} \\
            SPC & OOT & OOT \\
            \textbf{SEAL (Ours)} & OOT & OOT \\
            \hline
            \end{tabular}
            }
            \caption{Performance comparison results of SEAL and latent feature methods on AnalogGenie.}
            \label{subtable:latent}
        \end{subtable}
        \par\vspace{0.6em}
        \begin{subtable}{\linewidth}
            \centering
            \resizebox{\linewidth}{!}{%
            \begin{tabular}{lcc}
            \hline
            Framework & \multicolumn{2}{c}{AnalogGenie} \\
            \cline{2-3}
            & Accuracy & ROC-AUC \\
            \hline
            VGAE & OOM & OOM \\
            GraphSAGE & \textbf{\textcolor{red}{79.36\% ± 0.17\%}} & \textbf{\textcolor{red}{0.9296 ± 0.0020}} \\
            GCN+LRGA & OOT & OOT \\
            \textbf{SEAL (Ours)} & OOT & OOT \\
            \hline
            \end{tabular}
            }
            \caption{Performance comparison results of SEAL and other GNN methods on AnalogGenie.}
            \label{subtable:gnn}
        \end{subtable}
    \end{minipage}
\end{table}

% \begin{figure}
%   \centering
%   \begin{minipage}{\textwidth}
%     \centering
%     \begin{subfigure}{0.49\linewidth}
%       \centering
%       \includegraphics[width=\linewidth]{scaling_analysis_with_fit.pdf}
%       \caption{Relationship between training time, RAM, VRAM and the average number of edges per graph.}
%     \end{subfigure}
%     \hfill
%     \begin{subfigure}{0.49\linewidth}
%       \centering
%       \includegraphics[width=\linewidth]{scaling_analysis_with_oot_point.pdf}
%       \caption{Relationship between training time, RAM, VRAM and the total number of edges in the dataset. (CHANGE TEXT!!)}
%     \end{subfigure}
%     \caption{Relationship between Time, RAM, VRAM and edges.}
%     \label{computational_complexity_relationship}
%   \end{minipage}
% \end{figure}

\begin{figure}
  \centering
  \begin{minipage}{0.48\textwidth}
    \centering
    \begin{subfigure}{\linewidth}
      \centering
      \includegraphics[width=\linewidth]{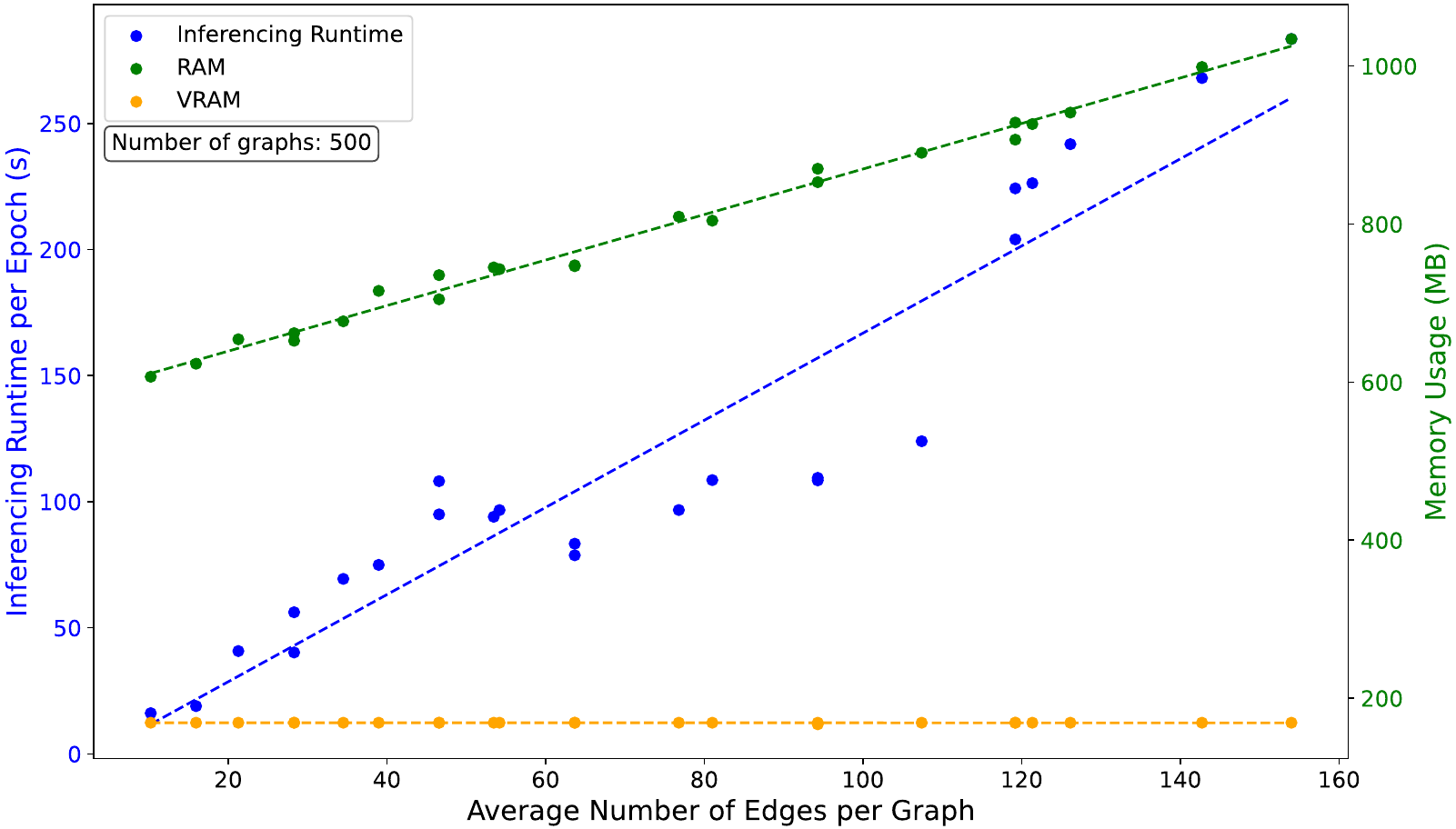}
      \caption{Relationship between inference time, RAM, VRAM and the average number of edges per graph.}
    \end{subfigure}
    \par\bigskip
    \begin{subfigure}{\linewidth}
      \centering
      \includegraphics[width=\linewidth]{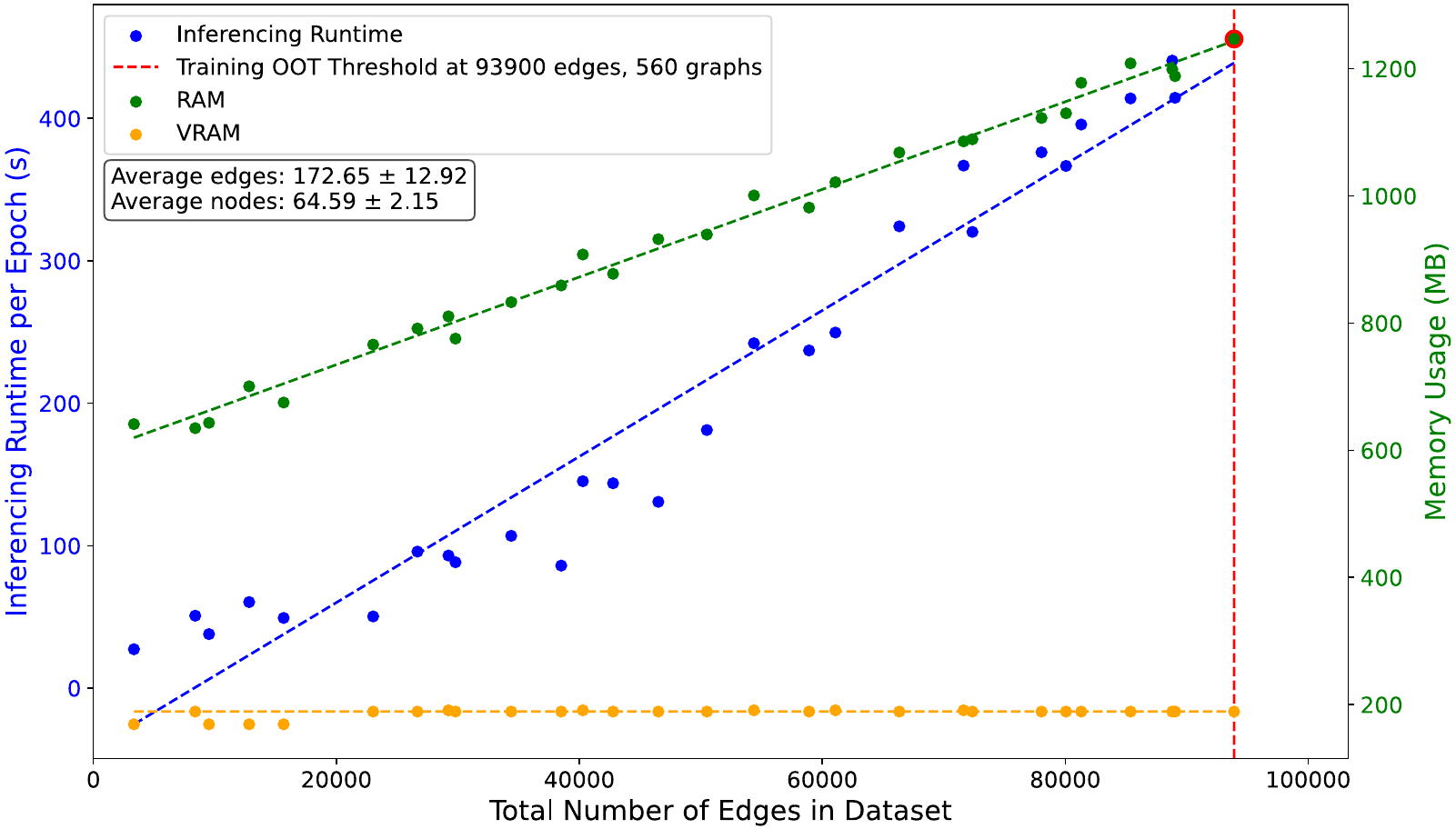}
      \caption{Relationship between inference time, RAM, VRAM and the total number of edges in the dataset.}
    \end{subfigure}
    \par\medskip
    \caption{Computational complexity analysis of SEAL. Inference time is measured in seconds.}
    \label{computational_complexity_relationship}
  \end{minipage}
  \hfill
  \begin{minipage}{0.48\textwidth}
    \centering
    \includegraphics[width=\linewidth]{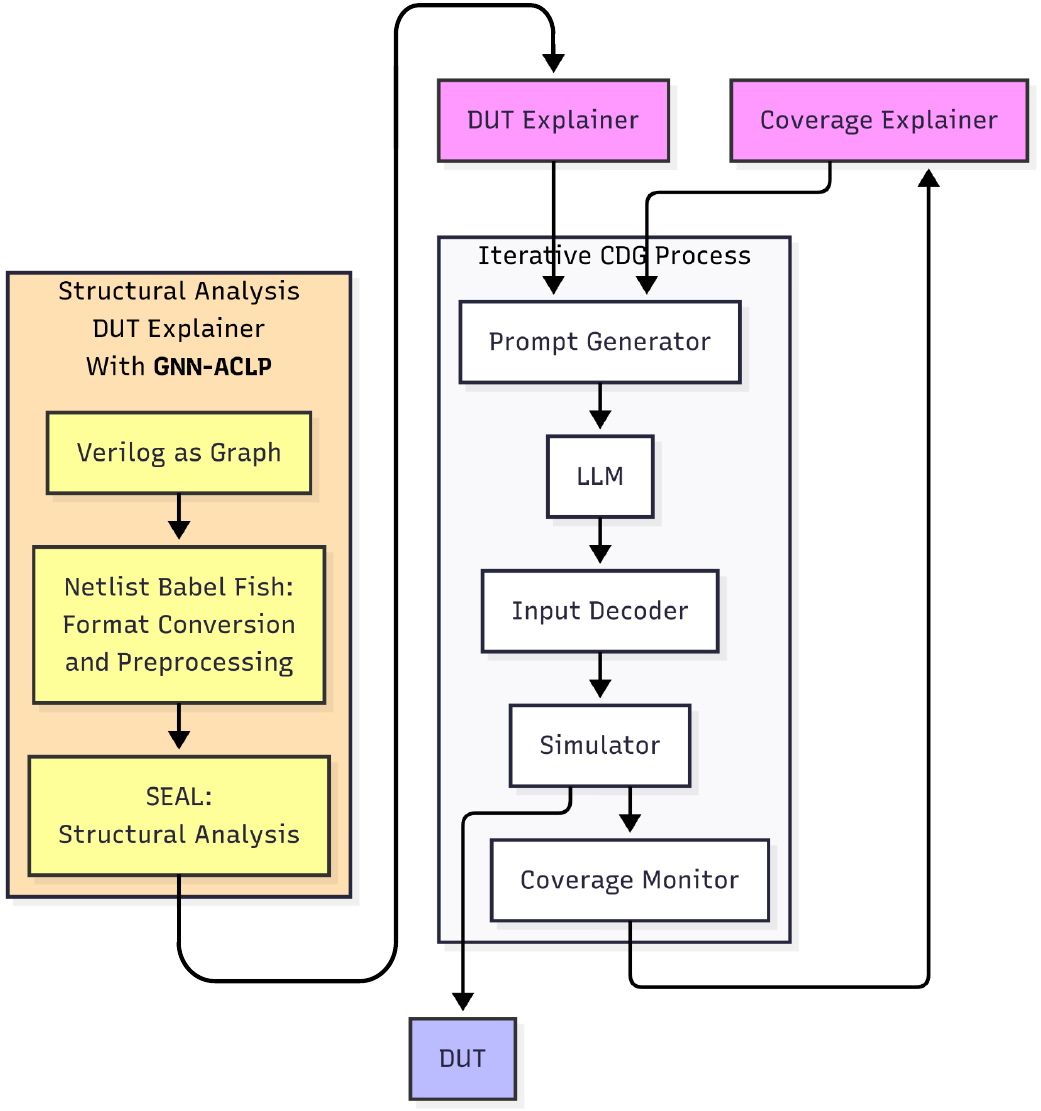}
    \caption{GNN-ACLP for structural analysis in VerilogReader.}
    \label{verilogreader}
    \centering
    \par\medskip
    \captionof{table}{Performance comparison results of the baseline, ScaLed, SUREL+ and SEAL on SpiceNetlist. The bold \textbf{\textcolor{red}{red}} and \textbf{\textcolor{blue}{blue}} fonts denote the best and the sub-optimal results,~respectively.}
    \label{discussion2}
    \resizebox{\linewidth}{!}{%
        \begin{tabular}{lccc}
        \hline
        Framework & Accuracy & ROC-AUC & Avg. Inf Time (s) \\
        \hline
        Baseline & 77.96\% ± 1.71\% & 0.8774 ± 0.0056 & \textbf{\textcolor{blue}{0.0640 ± 0.0045}} \\
        ScaLed & \textbf{\textcolor{blue}{82.88\% ± 2.69\%}} & 0.9136 ± 0.0312 & 0.2700 ± 0.0098 \\
        SUREL+ & 70.80\% ± 0.04\% & \textbf{\textcolor{blue}{0.9351 ± 0.0001}} & \textbf{\textcolor{red}{0.0023 ± 0.0002}} \\
        \textbf{SEAL (Ours)} & \textbf{\textcolor{red}{94.04\% ± 0.70\%}} & \textbf{\textcolor{red}{0.9866 ± 0.0019}} & 0.0836 ± 0.0069 \\
        \hline
        \end{tabular}}%
  \end{minipage}
\end{figure}

% \begin{figure}
%   \centering
%   \begin{minipage}{0.76\textwidth}
%     \centering
%     \begin{subfigure}{\linewidth}
%       \centering
%       \includegraphics[width=\linewidth]{103029_2.pdf}
%     \end{subfigure}
%   \end{minipage}
%   \par\medskip
%   \caption{Illustration of GNN-ACLP for structural analysis in VerilogReader.}
%   \label{verilogreader}
% \end{figure}

We attribute the observed OOT issue in GNN-ACLP for AnalogGenie to the inherent computational complexity of SEAL, particularly due to its node labeling approach \cite{Zhang2022}. Notably, SEAL’s performance tends to degrade when applied to highly dense graph structures since its labeling approach struggles to distinguish critical circuit subgraphs \cite{zhang2020revisiting}, a key challenge in AnalogGenie’s highly dense networks.

We further investigate the computational complexity of SEAL. Specifically, we conduct experiments to determine the relationship between inference time\footnote{Inference time includes the time for subgraph extraction, encoding and GNN inference on all graphs of the test dataset.}, RAM, VRAM, the average number of edges per graph, and the total number of edges in the dataset. We demonstrate the results in Figure \ref{computational_complexity_relationship}. Due to the subgraph extraction nature of SEAL, the usage of VRAM remains essentially constant. However, runtime, RAM, and VRAM increase linearly with both the average number of edges per graph and the total number of edges in the dataset, suggesting a linear computational complexity.

To address these challenges, we will optimize SEAL’s enclosing subgraph extraction by filtering electrically irrelevant nodes \cite{golovach2025finding} and introduce a circuit-specific labeling approach to reduce complexity without sacrificing accuracy in the future. Furthermore, we investigate alternative subgraph-based methods that more effectively represent structures in dense graphs, thereby addressing the limitations of SEAL in such scenarios. Specifically, we compare the accuracy, ROC-AUC, and average inference time per graph\footnote{We define the average inference time as the average time for subgraph extraction, encoding, and GNN inference across all graphs of the test dataset.} of the baseline, SEAL, ScaLed \cite{louis2022sampling} and SUREL+ \cite{yin2023surel+} on SpiceNetlist. ScaLed samples sparse enclosing subgraphs using a sequence of random walks, and SUREL+ samples node sets offline and joins them online as a proxy for a subgraph. We show the results in Table \ref{discussion2}. In our experiments with SpiceNetlist, SEAL demonstrates a moderate average inference time per graph, whereas SUREL+ achieves a lower average inference time per graph. However, the accuracy and ROC-AUC metrics for ScaLed and SUREL+ are significantly lower than those for SEAL in these experiments, indicating that lighter subgraph extraction methods such as ScaLed and SUREL+ may compromise prediction performance compared to SEAL.

\section{Future Work}

We believe GNN-ACLP will serve as a pivotal component of an end-to-end pipeline for analog circuit design automation.

\paragraph{\textbf{GNN-ACLP for Structural Verification and Consistency Constraints}}
Recent research has highlighted the potential of adapting LLMs for analog circuit design automation \cite{fu2023gpt4aigchip_1, chipnemo2023_2, yin2024ado_30, liu2024ladac_31}. However, LLMs exhibit limitations in precise mathematical computations \cite{ahn-etal-2024-large_3}, relational understanding \cite{li2024llms_4}, and logical reasoning \cite{giadikiaroglou-etal-2024-puzzle_5}. They often require manual instructions and verification \cite{fang2025survey_6} and are prone to generating incorrect outputs confidently, a phenomenon known as hallucination \cite{li2025circuit_1000}. In this context, GNN-ACLP can serve as a structural verification and repair module, automatically detecting and correcting possible missing or incorrect connections in preliminary netlists generated by LLMs. Beyond working with LLM, GNN-ACLP's structural verification capabilities can also benefit broader circuit design workflows. It can serve as a structural consistency constraint module within unified framework methods \cite{zhao2022analog_98}, ensuring that generated circuit topologies maintain structural validity.

Consider VerilogReader \cite{ma2024verilogreader} integrating LLM into the Coverage Directed Test Generation (CDG) process to aid hardware test generation. In this context, the LLM acts as a Verilog reader. Due to its complex structure, characterized by multiple concurrent always blocks and intricate module hierarchies, Verilog presents a significant challenge for LLMs to interpret directly. To address this issue, employing GNN-ACLP for structural analysis can enhance the scalability of AI's understanding of hardware. As demonstrated in Figure \ref{verilogreader}, the process begins with parsing the Verilog Design Under Test (DUT) to convert the RTL code into graphs. The Netlist Babel Fish then corrects errors and converts the data into a standardized JSON format. Subsequently, SEAL learns embeddings and extracts structural insights, such as module connectivity patterns, hard-to-reach states in finite state machines (FSMs), and data flow graphs, and provides a structural description that enhances the DUT Explainer's inputs.

% \begin{figure}
%   \centering
%   \begin{minipage}{0.76\textwidth}
%     \centering
%     \begin{subfigure}{\linewidth}
%       \centering
%       \includegraphics[width=\linewidth]{103029_2.pdf}
%     \end{subfigure}
%   \end{minipage}
%   \par\medskip
%   \caption{Illustration of GNN-ACLP for structural analysis in VerilogReader.}
%   \label{verilogreader}
% \end{figure}

\paragraph{\textbf{Netlist Babel Fish for Cross-Format Netlist Unification}}
As previously stated, the diversity of netlist formats poses significant challenges in analog design automation, substantially increasing maintenance costs \cite{vinagrero2023python}. For instance, a critical obstacle lies in identifying the functional blocks that are fundamental to constructing hierarchical netlist representations. However, the diversity of netlist formats compromises the consistency of topological structure representations and reduces the generalizability of computational models \cite{kunal2023gnn}. Given these constraints, Netlist Babel Fish leveraging RAG and LLMs can mitigate these interoperability barriers and enable the unification of cross-format netlists.

\paragraph{\textbf{GNN-ACLP for Analog Circuit Design Automation Explainability and Interpretability}}
The integration of trustworthy algorithms with explainability and interpretability is critically important for analog circuit design automation \cite{krishnamurthy2020explaining}. The SEAL framework within GNN-ACLP, as a subgraph-based GNN method, provides explicit explanations for predicted edges and demonstrates superior interpretability. Furthermore, Netlist Babel Fish generates natural language explanations while cross-referencing multiple netlist formats, thereby improving transparency, usability, and engineer trust.

Consider a case where a bandgap reference circuit experiences a post-simulation failure. Although passing the DC simulation, it fails the transient startup simulation because the output voltage does not stabilize correctly. In situations like this, an engineer may have to painstakingly trace signals, verify device sizes and models, scrutinize biasing, and hypothesize about potential missing or incorrect connections. This iterative debugging process can involve multiple simulation runs with incremental changes, often resulting in a time-consuming loop that could take hours or even days. By passing the corresponding netlist to GNN-ACLP, which learns patterns from correctly designed bandgap circuits, it can identify connections with a low probability of existence and provide explainability and interpretability for the engineer to focus their investigation on specific connections.

\section{Conclusion}

We propose GNN-ACLP, a novel graph neural networks based framework with SEAL framework integration, achieving port-level accuracy in analog circuit link prediction. We also propose Netlist Babel Fish, a netlist format conversion tool combining RAG with an LLM. Additionally, we create the SpiceNetlist dataset for training and evaluation of circuit link prediction methods. The results of the experiment demonstrate that our method achieves state-of-the-art performance. However, its linear computational complexity makes processing large-scale netlists challenging and requires future addressing.

\printcredits

\section*{Declaration of competing interest}

This research was supported by National Natural Science Foundation of China under Grant No. 62571173 and No. 62201323, Zhejiang Provincial Natural Science Foundation of China under Grant No. LD25F020005, No. LDT23F01015F01 and No. LQN25F030009.

\bibliographystyle{elsarticle-num}

\bibliography{merged_references,supp,supp2,supp3}

\clearpage
\appendix

\section{Extensibility of Netlist Babel Fish}
\label{extensibility_NBF}%%%

In the extensibility experiments for Netlist Babel Fish, we replace examples of SPICE and JSON netlists with examples of SPICE, Verilog-A, and VHDL-AMS netlists. We utilize DeepSeek-V3.1 \cite{deepseekDeepSeekV31} as the LLM here. Then, we test three examples, and Netlist Babel Fish converts all of them correctly.

\setminted{breaklines}

\paragraph{\textbf{Example 1}}

Original SPICE netlist:
\begin{minted}{SPICE}
.title KiCad schematic
C1 Net-_C1-Pad1_ NC_01 1000p
L1 Output Net-_C1-Pad1_ 10u
R1 Output GND 330
.end
\end{minted}
Converted Verilog-A netlist:
\begin{minted}{Verilog}
`include "disciplines.vams"

module circuit (Output, GND);
    inout Output, GND;
    electrical Output, GND;
    electrical Net_C1_Pad1, NC_01;

    parameter real C1_value = 1000e-12;
    parameter real L1_value = 10e-6;
    parameter real R1_value = 330;

    C_C1 (Net_C1_Pad1, NC_01) capacitor(C=C1_value);
    L_L1 (Output, Net_C1_Pad1) inductor(L=L1_value);
    R_R1 (Output, GND) resistor(R=R1_value);
endmodule
\end{minted}
Converted VHDL-AMS netlist:
\begin{minted}{VHDL}
library IEEE;
use IEEE.math_real.all;
use IEEE.electrical_systems.all;

entity circuit is
    port (
        terminal Output, GND : electrical
    );
end entity circuit;

architecture behavior of circuit is
    terminal Net_C1_Pad1, NC_01 : electrical;
    
    constant C1_value : real := 1000e-12;
    constant L1_value : real := 10e-6;
    constant R1_value : real := 330.0;
    
    quantity v_C1 across i_C1 through Net_C1_Pad1 to NC_01;
    quantity v_L1 across i_L1 through Output to Net_C1_Pad1;
    quantity v_R1 across i_R1 through Output to GND;
begin
    i_C1 == C1_value * v_C1'dot;
    v_L1 == L1_value * i_L1'dot;
    v_R1 == R1_value * i_R1;
end architecture behavior;
\end{minted}

%%%%%%

\paragraph{\textbf{Example 2}}

Original SPICE netlist:
\begin{minted}{SPICE}
*SPICE Netlist for circuit 806
I1 4 0 DC 1mA
I2 3 0 DC 1mA
R1 1 VDD 1k
M2 1 7 4 4 NMOS W=1u L=1u
R2 2 VDD 1k
M1 2 8 3 3 NMOS W=1u L=1u
R3 4 3 1k
.MODEL NMOS NMOS (LEVEL=1 VTO=1 KP=1.0e-4 LAMBDA=0.02)

.OP
.END
\end{minted}
Converted Verilog-A netlist:
\begin{minted}{Verilog}
`include "disciplines.vams"

module circuit_806 (vdd, gnd);
input vdd, gnd;
electrical vdd, gnd;

electrical n1, n2, n3, n4, n7, n8;

parameter real VDD = 5.0;

M2: D=n1, G=n7, S=n4, B=n4
M_M2 (n1, n7, n4, n4) nmos
(W=1u L=1u VTO=1.0 KP=1.0e-4 LAMBDA=0.02);

M1: D=n2, G=n8, S=n3, B=n3
M_M1 (n2, n8, n3, n3) nmos
(W=1u L=1u VTO=1.0 KP=1.0e-4 LAMBDA=0.02);

R1: n1 -> vdd, 1k
R_R1 (n1, vdd) resistor(R=1k);

R2: n2 -> vdd, 1k
R_R2 (n2, vdd) resistor(R=1k);

R3: n4 -> n3, 1k
R_R3 (n4, n3) resistor(R=1k);

I1: gnd -> n4, 1mA
I_I1 (gnd, n4) isource(DC=1m);

I2: gnd -> n3, 1mA
I_I2 (gnd, n3) isource(DC=1m);

V(vdd, gnd) <+ VDD;
endmodule
\end{minted}
Converted VHDL-AMS netlist:
\begin{minted}{VHDL}
library IEEE;
use IEEE.math_real.all;
use IEEE.electrical_systems.all;

entity circuit_806 is
    port (
        terminal vdd, gnd : electrical
    );
end entity circuit_806;

architecture behavior of circuit_806 is
    terminal n1, n2, n3, n4, n7, n8 : electrical;
    
    constant VTO : real := 1.0;
    constant KP : real := 1.0e-4;
    constant LAMBDA : real := 0.02;
    constant W : real := 1.0e-6;
    constant L : real := 1.0e-6;
    constant VDD_value : real := 5.0;
    
    quantity Vgs1 across Ig1 through n8 to n3;
    quantity Vds1 across Ids1 through n2 to n3;
    quantity Vgs2 across Ig2 through n7 to n4;
    quantity Vds2 across Ids2 through n1 to n4;
    
    quantity I_R1 through n1 to vdd;
    quantity I_R2 through n2 to vdd;
    quantity I_R3 through n4 to n3;
    quantity I_I1 through gnd to n4;
    quantity I_I2 through gnd to n3;
    
begin
    V(vdd, gnd) <= VDD_value;
    
    -- NMOS M1
    Ids1 == KP * (W / L) * ((Vgs1 - VTO) * Vds1 - 0.5 * Vds1**2) * (1.0 + LAMBDA * Vds1);
    
    -- NMOS M2
    Ids2 == KP * (W / L) * ((Vgs2 - VTO) * Vds2 - 0.5 * Vds2**2) * (1.0 + LAMBDA * Vds2);
    
    -- Resistor R1
    V(n1, vdd) == I_R1 * 1000.0;
    
    -- Resistor R2
    V(n2, vdd) == I_R2 * 1000.0;
    
    -- Resistor R3
    V(n4, n3) == I_R3 * 1000.0;
    
    -- Current Source I1
    I_I1 == 0.001;
    
    -- Current Source I2
    I_I2 == 0.001;
    
end architecture behavior;
\end{minted}

%%%%%%

\paragraph{\textbf{Example 3}}

Original SPICE netlist:
\begin{minted}{SPICE}
.title KiCad schematic
R2 RPI_19 3.3V 10k
Q1 D7 3.3V RPI_19 BS170
R1 D7 5V 10k
R4 RPI_18 3.3V 10k
Q2 D6 3.3V RPI_18 BS170
R3 D6 5V 10k
R6 RPI_17 3.3V 10k
Q3 D5 3.3V RPI_17 BS170
R5 D5 5V 10k
R8 RPI_16 3.3V 10k
Q4 D4 3.3V RPI_16 BS170
R7 D4 5V 10k
R10 RPI_15 3.3V 10k
Q5 D3 3.3V RPI_15 BS170
R9 D3 5V 10k
R12 RPI_14 3.3V 10k
Q6 D2 3.3V RPI_14 BS170
R11 D2 5V 10k
R14 RPI_13 3.3V 10k
Q7 D1 3.3V RPI_13 BS170
R13 D1 5V 10k
R16 RPI_12 3.3V 10k
Q8 D0 3.3V RPI_12 BS170
R15 D0 5V 10k
R18 RPI_10 3.3V 10k
Q9 RPI_SEL 3.3V RPI_10 BS170
R17 RPI_SEL 5V 10k
.end
\end{minted}
Converted Verilog-A netlist:
\begin{minted}{Verilog}
`include "disciplines.vams"

module circuit (D0, D1, D2, D3, D4, D5, D6, D7, RPI_SEL, RPI_10, RPI_12, RPI_13, RPI_14, RPI_15, RPI_16, RPI_17, RPI_18, RPI_19, V3_3V, V5V);
inout D0, D1, D2, D3, D4, D5, D6, D7, RPI_SEL, RPI_10, RPI_12, RPI_13, RPI_14, RPI_15, RPI_16, RPI_17, RPI_18, RPI_19, V3_3V, V5V;
electrical D0, D1, D2, D3, D4, D5, D6, D7, RPI_SEL, RPI_10, RPI_12, RPI_13, RPI_14, RPI_15, RPI_16, RPI_17, RPI_18, RPI_19, V3_3V, V5V;

NPN Q1 (D7, V3_3V, RPI_19);
NPN Q2 (D6, V3_3V, RPI_18);
NPN Q3 (D5, V3_3V, RPI_17);
NPN Q4 (D4, V3_3V, RPI_16);
NPN Q5 (D3, V3_3V, RPI_15);
NPN Q6 (D2, V3_3V, RPI_14);
NPN Q7 (D1, V3_3V, RPI_13);
NPN Q8 (D0, V3_3V, RPI_12);
NPN Q9 (RPI_SEL, V3_3V, RPI_10);

Resistor R1 (D7, V5V, 10k);
Resistor R2 (RPI_19, V3_3V, 10k);
Resistor R3 (D6, V5V, 10k);
Resistor R4 (RPI_18, V3_3V, 10k);
Resistor R5 (D5, V5V, 10k);
Resistor R6 (RPI_17, V3_3V, 10k);
Resistor R7 (D4, V5V, 10k);
Resistor R8 (RPI_16, V3_3V, 10k);
Resistor R9 (D3, V5V, 10k);
Resistor R10 (RPI_15, V3_3V, 10k);
Resistor R11 (D2, V5V, 10k);
Resistor R12 (RPI_14, V3_3V, 10k);
Resistor R13 (D1, V5V, 10k);
Resistor R14 (RPI_13, V3_3V, 10k);
Resistor R15 (D0, V5V, 10k);
Resistor R16 (RPI_12, V3_3V, 10k);
Resistor R17 (RPI_SEL, V5V, 10k);
Resistor R18 (RPI_10, V3_3V, 10k);

endmodule
\end{minted}
Converted VHDL-AMS netlist:
\begin{minted}{VHDL}
library IEEE;
use IEEE.math_real.all;
use IEEE.electrical_systems.all;

entity circuit is
    port (
        terminal D0, D1, D2, D3, D4, D5, D6, D7, RPI_SEL, RPI_10, RPI_12, RPI_13, RPI_14, RPI_15, RPI_16, RPI_17, RPI_18, RPI_19, V3_3V, V5V : electrical
    );
end entity circuit;

architecture behavior of circuit is
    component NPN is
        generic (R : real := 1.0e3);
        port (terminal Base, Emitter, Collector : electrical);
    end component;
    
    component Resistor is
        generic (R : real := 1.0e3);
        port (terminal Pos, Neg : electrical);
    end component;

begin
    Q1: NPN generic map (R => 10.0e3) port map (D7, V3_3V, RPI_19);
    Q2: NPN generic map (R => 10.0e3) port map (D6, V3_3V, RPI_18);
    Q3: NPN generic map (R => 10.0e3) port map (D5, V3_3V, RPI_17);
    Q4: NPN generic map (R => 10.0e3) port map (D4, V3_3V, RPI_16);
    Q5: NPN generic map (R => 10.0e3) port map (D3, V3_3V, RPI_15);
    Q6: NPN generic map (R => 10.0e3) port map (D2, V3_3V, RPI_14);
    Q7: NPN generic map (R => 10.0e3) port map (D1, V3_3V, RPI_13);
    Q8: NPN generic map (R => 10.0e3) port map (D0, V3_3V, RPI_12);
    Q9: NPN generic map (R => 10.0e3) port map (RPI_SEL, V3_3V, RPI_10);

    R1: Resistor generic map (R => 10.0e3) port map (D7, V5V);
    R2: Resistor generic map (R => 10.0e3) port map (RPI_19, V3_3V);
    R3: Resistor generic map (R => 10.0e3) port map (D6, V5V);
    R4: Resistor generic map (R => 10.0e3) port map (RPI_18, V3_3V);
    R5: Resistor generic map (R => 10.0e3) port map (D5, V5V);
    R6: Resistor generic map (R => 10.0e3) port map (RPI_17, V3_3V);
    R7: Resistor generic map (R => 10.0e3) port map (D4, V5V);
    R8: Resistor generic map (R => 10.0e3) port map (RPI_16, V3_3V);
    R9: Resistor generic map (R => 10.0e3) port map (D3, V5V);
    R10: Resistor generic map (R => 10.0e3) port map (RPI_15, V3_3V);
    R11: Resistor generic map (R => 10.0e3) port map (D2, V5V);
    R12: Resistor generic map (R => 10.0e3) port map (RPI_14, V3_3V);
    R13: Resistor generic map (R => 10.0e3) port map (D1, V5V);
    R14: Resistor generic map (R => 10.0e3) port map (RPI_13, V3_3V);
    R15: Resistor generic map (R => 10.0e3) port map (D0, V5V);
    R16: Resistor generic map (R => 10.0e3) port map (RPI_12, V3_3V);
    R17: Resistor generic map (R => 10.0e3) port map (RPI_SEL, V5V);
    R18: Resistor generic map (R => 10.0e3) port map (RPI_10, V3_3V);

end architecture behavior;
\end{minted}

\end{document}